\documentclass{aa}
\usepackage{graphicx}
\usepackage{txfonts}
\usepackage{multirow}
\usepackage{CJKutf8}
\usepackage{rotating}
\usepackage{afterpage}
\usepackage{changepage}
\usepackage{sidecap}
\usepackage{lipsum}
\usepackage{subcaption}
\usepackage{lscape}
\usepackage{placeins}
\usepackage{booktabs}
\usepackage[dvipsnames]{xcolor}
\usepackage[colorlinks=true, citecolor=blue]{hyperref}
\usepackage{stfloats}

\newcommand{\lmfitfull}{\textit{Non-Linear Least-Squares Minimization and Curve-Fitting}}
\newcommand{\ppxf}{\texttt{pPXF}}
\newcommand{\sps}{stellar population synthesis}
\newcommand{\jwst}{JWST}
\newcommand{\nirspec}{NIRSpec}
\newcommand{\DD}{\textit{DeepDive}}

\newcommand{\lmfit}{\texttt{LMFIT}}
\newcommand{\reff}{\ensuremath{r_\mathrm{eff}}\xspace}

\newcommand{\dMout}{$\log(\dot M_{\rm out} / \mathrm{M_\odot \, yr}^{-1})$}

\newcommand{\voff}{$\Delta v$}
\newcommand{\vout}{$\mathrm{v_{out}}$}
\newcommand{\Msunyr}{$\rm M_{\odot}\,\,\mathrm{yr^{-1}}$}
\newcommand{\kms}{$\mathrm{km\,s^{-1}}$}

\defcitealias{sunEvolutionGasFlows2024}{SLZ24}
\defcitealias{daviesJWSTRevealsWidespread2024}{DBP24}

\def\nii{[\ion{N}{ii}]}
\def\oiii{[\ion{O}{iii}]}
\def\hi{\ion{H}{i}}
\def\hii{\ion{H}{ii}}
\def\halpha{H$\alpha$}
\def\hbeta{H$\beta$}
\def\nai{\ion{Na}{i}}

\def\sii{[\ion{S}{ii}]}
\newcommand{\mgii}{\ion{Mg}{ii}} 

\newcommand{\feii}{\ion{Fe}{ii}}
\newcommand{\hei}{\ion{He}{i}}

\newcommand{\zh}[1]{\begin{CJK}{UTF8}{bsmi}#1\end{CJK}}
\newcommand{\jp}[1]{\begin{CJK}{UTF8}{min}#1\end{CJK}}

\begin{document} 

   \title{There and back again? Neutral outflows in $z\sim3.5$ quiescent galaxies}

    \author{Pengpei Zhu (\zh{朱芃佩})\inst{1,2,3}\orcid{0000-0002-6768-8335}
    \and Kei Ito (\jp{伊藤慧})\inst{1,2}\orcid{0000-0002-9453-0381}
    \and Francesco Valentino \inst{1,2}\orcid{0000-0001-6477-4011}
    \and Massissilia Hamadouche \inst{4}\orcid{0000-0001-6763-5551}
    \and Gianluca Scarpe \inst{1,2}\orcid{0009-0001-9808-3419}
    \and Katherine E. Whitaker \inst{4,1}\orcid{0000-0001-7160-3632}
    \and Takumi Kakimoto (\jp{柿元拓実})\inst{5,6}\orcid{0000-0003-2918-9890}
    \and William M. Baker \inst{7}\orcid{0000-0003-0215-1104}
    \and Anna R. Gallazzi \inst{3}\orcid{0000-0002-9656-1800}
    \and Steven Gillman \inst{1,2}\orcid{0000-0001-9885-4589}
    \and Rashmi Gottumukkala \inst{1,8} \orcid{0000-0003-0205-9826}
    \and Christian Kragh Jespersen\inst{9}\orcid{0000-0002-8896-6496}
    \and Minju Lee \inst{1,2}\orcid{0000-0002-2419-3068}
    \and Allison W. S. Man \inst{10}\orcid{0000-0003-2475-124X}
    \and Georgios Magdis \inst{1,2}\orcid{0000-0002-4872-2294}
    \and Masato Onodera \inst{5,11}\orcid{0000-0003-3228-7264}
    \and Rhythm Shimakawa \inst{12}\orcid{0000-0003-4442-2750}
    \and Aswin Vijayan \inst{13}\orcid{0000-0002-1905-4194}
    \and Po-Feng Wu (\zh{吳柏鋒}) \inst{14,15,16}\orcid{0000-0002-9665-0440}
            }            

    \institute{
    Cosmic Dawn Center (DAWN), Denmark
    \and
    DTU Space, Technical University of Denmark, Elektrovej 327, 2800 Kgs. Lyngby, Denmark
    \and
    INAF-Osservatorio Astrofisico di Arcetri, Largo Enrico Fermi 5, I-50125 Firenze, Italy
    \and
    Department of Astronomy, University of Massachusetts, Amherst, MA 01003, USA
    \and
    Department of Astronomical Science, The Graduate University for Advanced Studies, SOKENDAI, 2-21-1 Osawa, Mitaka, Tokyo 181-8588, Japan
    \and
    National Astronomical Observatory of Japan, 2-21-1 Osawa, Mitaka, Tokyo 181-8588, Japan
    \and
    DARK, Niels Bohr Institute, University of Copenhagen, Jagtvej 155A, DK-2200 Copenhagen, Denmark
    \and
    Niels Bohr Institute, University of Copenhagen, Jagtvej 128, DK-2200 Copenhagen N, Denmark
    \and
    Department of Astrophysical Sciences, Princeton University, Princeton, NJ 08544, USA
    \and
    Department of Physics \& Astronomy, University of British Columbia, 6224 Agricultural Road, Vancouver BC, V6T 1Z1, Canada
    \and
    Subaru Telescope, National Astronomical Observatory of Japan, National Institutes of Natural Sciences (NINS), 650 North A’ohoku Place, Hilo, HI 96720, USA
    \and
    Waseda Institute for Advanced Study (WIAS), Waseda University, 1-21-1, Nishi-Waseda, Shinjuku, Tokyo 169-0051, Japan
    \and
    Astronomy Centre, Department of Physics and Astronomy, Pevensey II Building, University of Sussex, Brighton BN1 9QH, UK
    \and
    Graduate Institute of Astrophysics, National Taiwan University, Taipei 10617, Taiwan
    \and
    Department of Physics and Center for Theoretical Physics, National Taiwan University, Taipei 10617, Taiwan
    \and
    Physics Division, National Center for Theoretical Sciences, Taipei 10617, Taiwan
    }

    \titlerunning{}
    \authorrunning{P. Zhu et al.}

   \date{Received February 19, 2026; accepted May 20, 2026}
 
    \abstract{Neutral gas outflows play a crucial role in the baryon cycle of galaxies, regulating their evolution by removing gas and redistributing energy and momentum into the surrounding medium. Their properties provide key insights into the transition from star formation to quiescence. In this work, we investigate the neutral gas outflow properties of 23 massive ($M_\star = 10^{10.1-11.6}\,\rm M_\odot$) quiescent galaxies (QGs) at $z=2.82$--$4.61$, selected from the JWST NIRSpec ($\rm R\sim1000$) spectroscopic and NIRCam imaging program \DD. We trace the neutral gas outflows using the \nai\ Doublet absorption lines and detect excess \nai~D absorption in 13/23 (57\%) targets, of which 7/23 (30\%) show blueshifted excess absorption with velocity offsets |\voff|$\gtrsim 150\,\,\mathrm{km\,s^{-1}}$. The $z\sim3.5$ targets exhibit velocity offsets similar to those of their local massive quiescent counterparts; they are also equivalent when compared in SFR--\voff\ space. We derive mass outflow rates and identify, in particular, the most extreme neutral gas outflow rate \dMout$=2.68\pm0.27$ ever reported beyond the local Universe, coincident with an X-ray AGN. For all \nai~D detected systems, the inferred mass outflow rate can, in principle, suppress ongoing star formation (i.e., $\mathrm{SFR}\leq\dot{M}_{\mathrm{out}}$); however, the outflows are unlikely to escape their hosts, and are suggestive of fountain-like recycling on relatively short timescales ($\sim3$--$180$ Myr), depending on the assumed potential and launching radius. All \nai~D detected targets occupy the LI(N)ER region of the BPT diagram and/or are X-ray detected. Still, we find no strong correlation between ongoing AGN activity and the neutral outflow: 2/4 broad-line/X-ray AGNs are \nai~D undetected -- yet, the outflows can be powered by fossil/episodic AGNs, and one broad-line target shows a possible P-Cygni profile that indicates strong outflows. As, on average, neutral outflows alone are not able to permanently quench star formation by removing gas in our sample at $z\sim3.5$, the presence of gas cycling in and out of massive passive systems may instead be the signature of feedback-regulated quenching-maintenance processes.}


   \maketitle
   \nolinenumbers

\section{Introduction}
\label{sec:intro}

Understanding the mechanisms that quench star formation in massive galaxies is crucial to studies of galaxy evolution. Over the past decade, deep optical spectroscopy has revealed a new population of massive ($M_\star\sim10^{10-11}\,\mathrm{M_\odot}$) quiescent systems at $z>3$ \citep[e.g.,][]{glazebrookMassiveQuiescentGalaxy2017, schreiberInfraredSpectroscopyStarformation2018, tanakaStellarVelocityDispersion2019, valentinoQuiescentGalaxies152020, forrestExtremelyMassiveQuiescent2020, forrestMAGAZ3NEHighStellar2022, carnallMassiveQuiescentGalaxy2023, nanayakkaraPopulationFaintOld2024, settonLittleRedDots2025, degraaffEfficientFormationMassive2024, bakerAbundanceNatureHighredshift2025}, whose number density, early formation, and abrupt quenching remain difficult to reconcile with standard prescriptions in cosmological simulations \citep[e.g.,][]{schreiberInfraredSpectroscopyStarformation2018, merlinRedDeadCANDELS2019, valentinoQuiescentGalaxies152020, lagosDiverseStarFormation2025, bakerAbundanceNatureHighredshift2025, bakerExploring700Massive2025}. Nevertheless, the physical drivers of the transition from star-bursting to quiescent --- whether due to active galactic nuclei (AGN), stellar feedback, merging, or a combination thereof --- are still debated.

Gas outflows provide a direct probe of the AGN or stellar feedback mechanisms that can deplete or redistribute the galactic cold gas reservoir --- especially the molecular gas phase that directly fuels star formation --- thereby shutting down star formation and/or maintaining the quiescent state \citep[][]{ciconeMassiveMolecularOutflows2014, harrisonImpactSupermassiveBlack2017, manStarFormationQuenching2018, herrera-camusMolecularIonizedGas2019,veilleuxCoolOutflowsGalaxies2020, manExquisitelyDeepView2021}. In particular, observations show that molecular outflows can dominate the mass and energetics of the wind and are capable of efficiently depleting or dynamically disturbing the star-forming gas reservoir \citep[][]{ciconeMassiveMolecularOutflows2014, herrera-camusMolecularIonizedGas2019}. In the local Universe, multiphase outflows are well established across starburst, post-starburst, and quiescent populations \citep[e.g.,][]{heckmanAbsorptionLineProbesGas2000, rupkeOutflowsInfraredLuminousStarbursts2005, ciconeMassiveMolecularOutflows2014, baronMultiphaseOutflowsPoststarburst2020, baronMultiphaseOutflowsPoststarburst2022}. Neutral outflows traced by the \nai~$\lambda\lambda5891,5897\AA$ doublet, in particular, offer sensitivity to cool neutral gas that often dominates the mass budget of winds: in galaxies with both neutral and ionized outflows, the neutral outflow rates are typically 10–100 times larger with respect to the ionized outflow rates \citep[e.g., ][]{fioreAGNWindScaling2017, roberts-borsaniObservationalConstraintsMultiphase2020, averyCoolOutflowsMaNGA2022, baronMultiphaseOutflowsPoststarburst2022, belliStarFormationShut2024}, but such measurements are mostly limited to the local Universe. Large SDSS samples show that neutral outflows are common in massive star-forming and post-starburst galaxies, but weaken along the evolutionary sequence toward quiescence, and shift to net inflow for the most quiescent local galaxies \citep[][hereafter SLZ24]{concasTwofacesIonizedNeutral2019, sunEvolutionGasFlows2024}. These trends suggest a decline in outflow launching efficiency as star formation fades, with AGN activity contributing to the most energetic winds \citep{baronMultiphaseOutflowsPoststarburst2022}.

At higher redshifts ($z>1$), the landscape differs. Pre-\jwst\ studies were limited by wavelength coverage and sensitivity, probing neutral outflows mainly via UV tracers such as \mgii\ and \feii\; and these studies are largely based on stacked spectra that cannot easily trace correlations between outflow properties and other physical parameters \citep[e.g.,][]{weinerUBIQUITOUSOUTFLOWSDEEP22009, steidelStructureKinematicsCircumgalactic2010, bordoloiDependenceGalacticOutflows2014, maltbyHighvelocityOutflowsMassive2019, manExquisitelyDeepView2021}. \jwst/NIRSpec has now enabled direct detection of \nai-traced outflows well beyond the local Universe. Powerful neutral winds have been reported in rapidly quenching systems at $z\sim2-4$, where the neutral phase carries most of the outflowing mass \citep{belliStarFormationShut2024, deugenioFastrotatorPoststarburstGalaxy2024, daviesJWSTRevealsWidespread2024, parkWidespreadRapidQuenching2024, wuEjectiveFeedbackQuenching2025, liboniProbingNeutralOutflows2026, taylorJWSTEXCELSSurvey2026}. These studies typically find broad, blueshifted \nai\ absorption with velocities of $\sim100$ to $\sim1000$~km\,s$^{-1}$ and mass outflow rates large enough to impact or even dominate the host's baryon budget. Such winds are frequently associated with AGN signatures in emission-line ratios or broad components, implying a nuclear contribution to the launch mechanism \citep{deugenioFastrotatorPoststarburstGalaxy2024, daviesJWSTRevealsWidespread2024, parkWidespreadRapidQuenching2024, taylorJWSTEXCELSSurvey2026}.
Beyond individual detections and small samples, a recent study \citep{lyuFirstStatisticalDetection2026} provides a statistical measurement of cool outflows in $M_\star>10^8\,\,\rm M_{\odot}$ across $1<z<10$ using archival \jwst\ stacked \mgii\ absorption, revealing ubiquitous blueshifted features even at $z>5$ and a nearly un-evolving velocity scale of $\sim$300~km\,s$^{-1}$. Their results demonstrate that cool outflows are widespread in typical high-redshift galaxies.

Yet, the connection between outflows and quenching is not straightforward. Some newly quenched galaxies exhibit relatively modest outflows, possibly residuals of the preceding starburst \citep{valentinoGasOutflowsTwo2025}. Conversely, extreme neutral outflows have been identified in galaxies lacking clear contemporaneous AGN activity, suggesting delayed or ``fossil'' winds \citep{sunExtremeNeutralOutflow2026}. The diversity uncovered at $z>2$ indicates that multiple pathways may give rise to neutral outflows, and that their role in quenching is sensitive to both timescales and gas-phase composition.

Simulations likewise predict that the bulk of outflowing mass resides in the neutral and molecular phases, but they diverge in the relative importance of stellar versus AGN feedback \citep[e.g.,][]{richingsOriginFastMolecular2018, nelsonFirstResultsTNG502019, hopkinsWhatCosmicRays2020, kimFirstResultsSMAUG2020}. Observational constraints on neutral outflows across cosmic time, therefore, provide essential tests of these feedback prescriptions.

In this context, our JWST Cycle 2 \DD\,(DD) program \citep[PID \#3567, PI. F. Valentino,][]{itoDeepDiveDeepDive2025a} provides a systematic census of \nai-traced neutral gas outflows in a sample of massive quiescent galaxies at $z\gtrsim3$. These galaxies belong to the earliest well-established quiescent population and probe the onset of the quenching era, when star formation must proceed both rapidly and efficiently, given the intrinsically short timescales of the early Universe. With medium-resolution ($R\sim1000$) \jwst/NIRSpec spectra, we measure the kinematics and strengths of \nai~D absorption to assess the incidence, velocity structure, and energetics of neutral outflows in these systems, and to compare them with both local analogs and recently quenched galaxies at intermediate redshift. In this work, we show that neutral outflows are common in massive QGs at $z>3$, with mass outflow rates comparable to or exceeding the ongoing star formation rate, and that they mostly cannot escape their hosts. By placing our sample in the broader context of outflow studies across cosmic time, we assess whether these winds can drive or maintain quiescence and evaluate the extent to which they are linked to (past) AGN activity.

The structure of this paper is as follows: In Section \ref{sec: methods}, we describe the \DD\ program, the size measurement, stellar continuum modeling, the excess \nai~D absorption model fitting method, and the derivation of outflow properties. In Section \ref{sec: results}, we present the results of the outflow properties, discuss their implications, and compare them with the literature. In Section \ref{sec: escape}, we apply simple models to estimate whether the outflows recycle or escape their hosts. In Section \ref{sec: AGN} we present the possibility of AGN presence in the \DD\ sample. In Section \ref{sec: discuss} we discuss the results of this paper, specifically about the SFR--outflow correlation, the driving mechanisms, and whether the outflow recycles, and if it is capable of quench/maintain the quenching of the host galaxy. Finally, in Section \ref{sec: summary} we summarize our work. In this work, we assume a flat Planck15 $\Lambda$CDM cosmology with \hbox{H$_0=67.74$ km s$^{-1}$ Mpc$^{-1}$}, $\Omega_m=0.3089$, and $\Omega_{\Lambda}=0.6911$ \citep{duttonColdDarkMatter2014}.

\section{Data and methods}
\label{sec: methods}

\subsection{The \DD~sample}
\label{sec: deepdive}

\DD~(DD) is a JWST General Observer program targeting quiescent galaxies at $z\sim3-4$ with NIRCam and NIRSpec. The NIRSpec observations were conducted with the G235M/F170LP grating to detect spectral features at rest-frame optical wavelengths. Ten primary targets are drawn from pre-JWST deep fields using extended $UVJ$ criteria \citep{williamsDetectionQuiescentGalaxies2009} to include recently quenched systems, while 17 secondary targets are selected based on multiple criteria ($UVJ$ colors, low sSFR, and $D_{\rm n}4000$). Details about the \DD\ sample selection and data reduction can be found in \citet{itoDeepDiveDeepDive2025a}, following the same methodology and using software tools ({\sc grizli}, \citealt{brammerGrizli2023}; {\sc msaexp}, \citealt{brammerMsaexpNIRSpecAnalyis2023}; \citealt{graaffRUBIESCompleteCensus2025}) adopted for the Dawn JWST Archive (DJA\footnote{\url{https://dawn-cph.github.io/dja/}}) data products.

\citet{hamadoucheDeepDiveTracingEarly2026} model the star formation histories (SFHs) of the \DD~targets with \texttt{Bagpipes} \citep{carnallInferringStarFormation2018}, and found that most of them recently ($\sim$400 Myr) concluded their major star-formation that lasted for a range of short (100-200 Myr) and long ($>300$ Myr) timescales, and now they have very low ongoing (on 100 Myr timescales) star formation rates (SFRs, $\sim 1$ \Msunyr).

We begin by examining the 10 \DD~primary targets + 17 \DD~secondary targets, for a total of 27 massive QGs. To estimate the excess \nai~D absorption from the ISM, careful removal of the stellar continuum contribution is required. We excluded DD-82 because it emits very strong Balmer lines that infill the stellar absorption features, making it difficult to separate the emission from the continuum and to constrain the stellar velocity dispersion from \sps\,(SPS). Moreover, some properties (strong emission lines, red and compact with \reff$=0.41^{+0.06}_{-0.05}$ kpc, shallow Balmer break) are reminiscent of those of a Little Red Dot \citep{mattheeLittleRedDots2024}, suggesting caution in interpreting its continuum emission. We also exclude DD-317 and DD-327 because their Balmer absorption lines are largely not covered by \nirspec, given their lower redshifts (z=2.68 and z=2.62, respectively), leaving us unable to constrain the SPS modeling. Finally, we exclude DD-144 as its \nai~D is partially blocked by the NIRSpec detection gap. After excluding DD-82, DD-144, DD-317, and DD-327 from further analysis, we obtain a final sample of 23 targets with $2.82 \leq z \leq 4.62$.

\subsection{Size measurement}
\label{sec: radius}
We measured the size (effective semi-major axis, $r_\mathrm{eff}$) of all targets with \texttt{pysersic} \citep{pashaPysersicPythonPackage2023}. A detailed description of the methods and tests for a complete photometric sample at $z=3-4$ is provided in Scarpe et al. (2026, in prep.). Here, we present the key points of their methods. We modeled the emission in the F200W band, broadly mapping the rest-frame optical wavelengths redward of the Balmer break at these redshifts. If unavailable, as for DD-106, DD-129, and DD-165, which are covered by the COSMOS-Web survey \citep{caseyCOSMOSWebOverviewJWST2023}, we modeled the surface brightness in the F277W band. For targets in major extragalactic fields, we retrieved NIRCam mosaics in DJA  (refer to \citealt{valentinoAtlasColorselectedQuiescent2023} for details). NIRCam/F200W images collected as part of the \DD\ program are reduced and analyzed consistently \citep{itoDeepDiveDeepDive2025a}.
We adopted a single S\'{e}rsic profile \citep{sersicInfluenceAtmosphericInstrumental1963} and masked potential contaminating sources within $3''\times 3''$ with a segmentation map generated with the Pythonic version of Source Extractor  \citep[\texttt{SEP};][]{bertinSExtractorSoftwareSource1996, barbarySEPSourceExtractor2016}. We used point spread functions (PSFs) from \citet{geninDAWNJWSTArchive2025} and \citet{itoDeepDiveDeepDive2025a} in the analysis. For consistency with \citet{itoMergingPairMassive2025}, we use $r_{\mathrm{eff}}$ measured in the F277W band for DD-196. In one case (DD-257), no NIRCam coverage at F200W or F277W is currently available; only $r_{\rm eff} = 2.12\pm0.04$ kpc in F356W is available, which is possibly smaller than its F200W size given the mild negative gradient of sizes with wavelength for quenched galaxies \citep[][Scarpe et al. 2026 in prep.]{bodanskyJWST+ALMARevealBuild2025}.

\begin{figure*}[b]
    \centering
    \includegraphics[width=\linewidth]{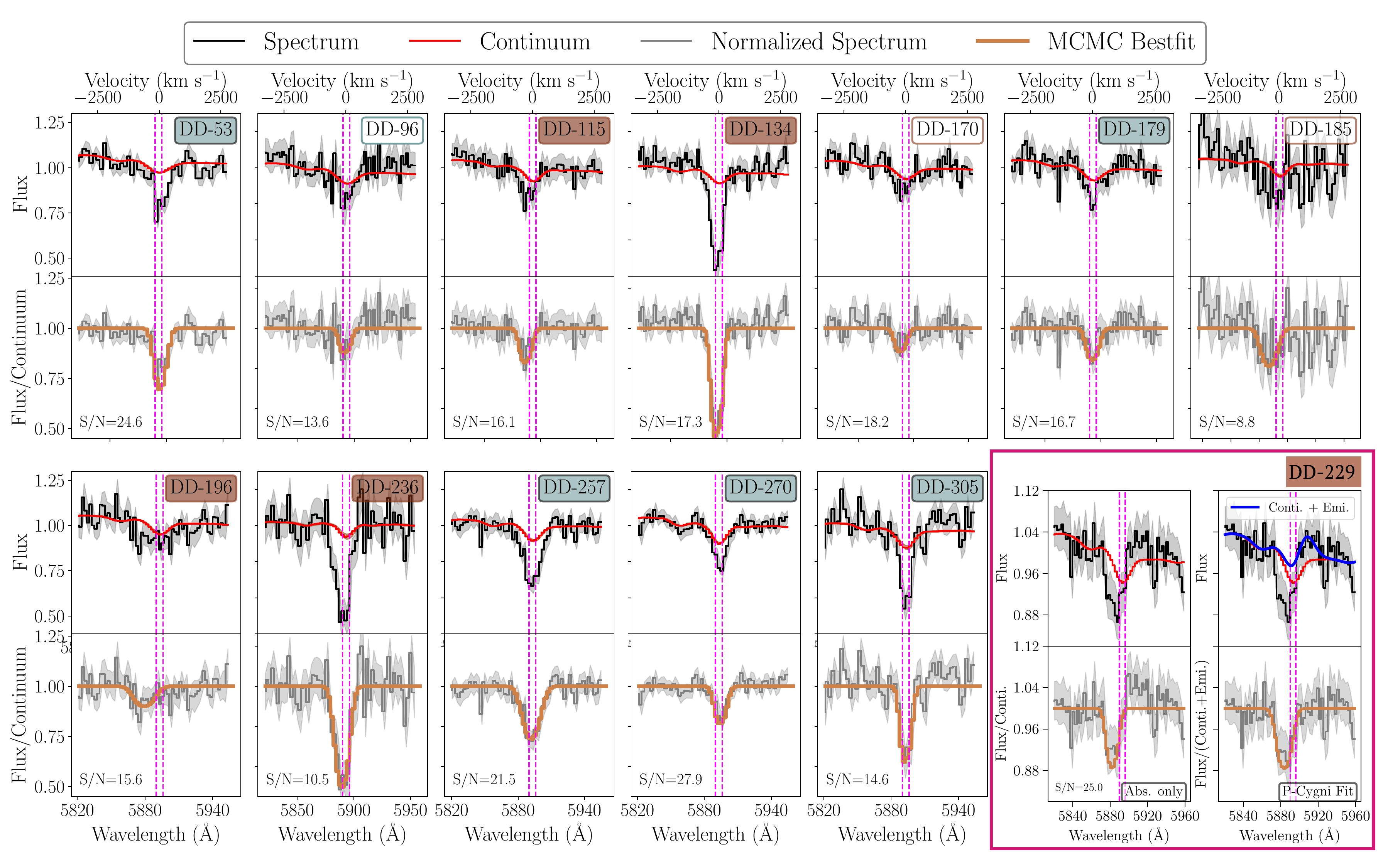}
    \caption{\DD~\nai~D detected spectra, including tentative detections. The first and third rows show the median-normalized spectra in black and best-fit stellar continuum in red; the second and fourth rows show the continuum-normalized (i.e., flux/continuum) spectra and the MCMC best-fit for \nai~D absorption. The median spectral S/N of each normalized spectrum is labeled on the lower left. The target's DD-ID is labeled on the upper right: the solid cyan and red labels represent 2-$\sigma$ detected systematic (or redshifted) and blueshifted excess \nai~D absorption, respectively; the hollow labels represent 1-$\sigma$ tentative detections (refer to Section \ref{sec: fitting} for detection definition). The two panels in the lower right show fittings with and without redshifted emission to fit the possible P-Cygni profile of DD-229. Additionally, DD-134 is ``Jekyll'' of the merging pair ``Jekyll \& Hyde'' \citep{schreiberJekyllHydeQuiescence2018}; DD-185 and DD-196 are reported by \citet{itoMergingPairMassive2025} as a merging pair with ids 61167 and 61168, respectively.}
    \label{fig:spec}
\end{figure*}

\subsection{Stellar population modeling}
In the context of this work, we aim to detect the interstellar \nai~D absorption (can be blueshifted/systemic/redshifted, hereafter excess \nai~D absorption), but there are also contributions to the galactic \nai~D absorption from stellar atmospheres \citep[e.g.,][]{alloinComparativeStudyNA1989, wortheyAbundanceRatioTrends1998}. Thus, careful removal of the stellar continuum is required in order to estimate the excess \nai~D absorption from gas. 

We modeled the stellar continuum over the wavelength range of $3700\AA-7200\AA$ using the penalized pixel fitting code \citep[\texttt{pPXF},][]{cappellariFullSpectrumFitting2023}, and focused on constraining the stellar velocity dispersion and depth of the stellar \nai~D absorption. We initialized \ppxf~by constraining stellar ages to be no older than 
the age of the Universe at the target redshift, and allow the stellar 
metallicity to range between $\mathrm{[M/H]}=-2$ and $0.5$.  
We used the Flexible Stellar Population Synthesis (FSPS) models \citep{conroyPropagationUncertaintiesStellar2010} together with the E-MILES stellar libraries \citep{vazdekisUVextendedEMILESStellar2016}. We followed the methodology applied in \citet{itoDynamicalPropertiesStar2026}, which consists of two steps. In the first step, we derived kinematic information, i.e., stellar velocity dispersion and velocity offset, the latter corresponding to a refined redshift estimate. Prior to fitting, spectra and stellar templates are flattened by modeling the spectral shape with a third-order B-spline function and dividing the spectra by it, thereby reducing the influence of the continuum shape. We masked spectra within a range of $\pm 40\AA$ from the wavelength of strong emission lines (\halpha, \hbeta, \oiii\,$\lambda\lambda 4959, 5007$, \nii\,$\lambda\lambda6548,6584$, and \sii\,$\lambda\lambda 6717,6731$) and a region encompassing the \nai~D (5780-5980 \AA) absorption, as they can affect the best-fit continuum. During the fitting, a second-order additive polynomial function was used to correct for imperfect background subtraction. This step was iterated 500 times, adding random noise with inverse-variance weights. In the second step, we ran \ppxf\ on the original, unflattened spectra, fixing the stellar velocity dispersion and velocity offset to the best-fit values from the first step. In this iteration, we masked only \nai~D (5780-5980 \AA), while emission lines are modeled as Gaussian functions. A second-order multiplicative polynomial function is used to correct for imperfect flux correction and to match its spectral shape to the stellar templates. This step was also iterated 500 times. The best-fit stellar templates and emission-line fluxes are determined from the median of the distribution obtained through the iterations.

\subsection{\nai~D models and fitting}

\subsubsection{Models}
In this work, we built two models for \nai~D with \lmfitfull~\citep[\lmfit,][]{newvilleLMFITNonLinearLeastSquares2025}: a simple two-Gaussian model and a partial covering model. The simple two-Gaussian model (Eq. \ref{eq: simple model}) is used to detect \nai~D using Monte-Carlo simulation (MCS), and to fit DD-229's possible P-Cygni emission profile following the method in \citet{baronMultiphaseOutflowsPoststarburst2022} (their Appendix B). The model operates in wavelength space and consists of two Gaussian components, allowing the  positive/negative amplitudes to be fit:

\begin{equation}
\label{eq: simple model}
F_{\mathrm{Na\,\textsc{i}\,D,\,simp.}}(\lambda)
= A_{D1} e^{-(\lambda-\lambda_{D1}\delta)^2/2\sigma_\lambda^2}
+ A_{D2} e^{-(\lambda-\lambda_{D2}\delta)^2/2\sigma_\lambda^2}.
\end{equation}

\noindent
where \(\lambda_{D1}=5897.6\,\text{\AA}\) and \(\lambda_{D2}=5891.6\,\text{\AA}\) are the wavelength of the \nai~D1/D2 lines. $\delta = 1+\frac{\Delta v}{c}$ is the Doppler factor. Amplitudes $A_{D1}, A_{D2}$ are positive for emission and negative for absorption, and the ratio $A_{D2}/A_{D1}$ varies between 1-2 for optically thick/thin cases.
\(\Delta v\) is the velocity offset of the line center, and $\sigma_\lambda$ is the line width in wavelength units.

For the absorption profile modeling, we used the partial covering model by \citet{rupkeOutflowsInfraredLuminousStarbursts2005}, which allows a direct comparison with recent studies of gas outflows based on JWST/NIRSpec spectra \citep{belliStarFormationShut2024, deugenioFastrotatorPoststarburstGalaxy2024, daviesJWSTRevealsWidespread2024, wuEjectiveFeedbackQuenching2025, liboniProbingNeutralOutflows2026, taylorJWSTEXCELSSurvey2026}. The model is given by:

\begin{equation}
\label{eq: physical model}
    F_{\mathrm{Na\,\textsc{i}\,D,\,abs.}}(v) = 1 - C_f + C_f \exp(-\tau_{D2}(\Delta v)-\tau_{D1}(\Delta v))
\end{equation}
\noindent
where $F_{\mathrm{Na\,\textsc{i}\,D,\,abs.}}$ is the excess \nai~D absorption on the continuum normalized flux, $C_f$ the covering factor varies from 0 to 1, and $\tau_{D2},\,\tau_{D1}$ are the optical depths of the \nai~D lines. For each line, we assume a Gaussian optical depth:

\begin{equation}
    \label{eq: tau}
    \tau(\Delta v,\sigma_v) = \tau_0\exp(-\Delta v^2/2\sigma_v^2)
\end{equation}
\noindent
We tied both lines in $\sigma_v$ and $\Delta v$, and fixed their line-center optical depth as $\tau_{0,\,D1} = 2\tau_{0,\,D2}$, consistent with the oscillator strength ratio \citepalias[see footnote 2 in][]{daviesJWSTRevealsWidespread2024}. To enable direct comparison with observed spectra, the analytic profile is first evaluated on a high-resolution wavelength grid and subsequently resampled to the instrumental resolution with {\sc msaexp} \citep{brammerMsaexpNIRSpecAnalyis2023}. In the calculation, we considered a 30\% higher resolution than the pre-launch nominal value \citep{graaffRUBIESCompleteCensus2025}. At the resolution of the G235M grating, the \nai~D lines are blended.

We fit the spectrum of each \DD\ QG between the wavelength region 5820 \AA\ to 5960 \AA, including contributions from the continuum ($F_{\star,\,\rm pPXF}$), the excess \nai~D absorption ($F_{\mathrm{Na\,\textsc{i}\,D,\,abs.}}$), and \nai~D emission ($F_{\mathrm{Na\,\textsc{i}\,D,\,emi.}}$ modeled by Eq. \ref{eq: simple model}, only for DD-229, following \citealt{baronMultiphaseOutflowsPoststarburst2022}):

\begin{equation}
\label{eq: full model}
    F(v) = (F_{\star,\,\rm pPXF} + F_{\mathrm{Na\,\textsc{i}\,D,\,emi.}}) \times  F_{\mathrm{Na\,\textsc{i}\,D,\,abs.}}
\end{equation}

We note the presence of \hei~$\lambda \,5876$ \AA\ emission in galactic spectra just blueward of the \nai~D line, typically originating from \hii\, regions or AGNs \citepalias{daviesJWSTRevealsWidespread2024}. However, none of the \nai~D (tentatively) detected \DD\ spectra show signs of \hei~emission; thus, we do not include \hei~as part of the modeling.

\subsubsection{Fitting}
\label{sec: fitting}

For both the simple and the partial-covering models, we first fit all 23 targets using least squares to obtain initial parameter estimates. We then adopt a Monte Carlo simulation (MCS)--based detection criterion, following \citet{zhuSQuIGGEObservationalEvidence2025}. For each target, we generate 3000 realizations of the continuum-normalized spectrum by perturbing the flux according to the per-pixel noise, and refit each realization using Eq.~\ref{eq: simple model}.This procedure yields a distribution of equivalent width (EW) measurements that naturally accounts for both noise and Gaussian line-profile assumptions. In each realization, the flux density is defined as $F_{\rm MCS}(\lambda) = F_{\star,\,\rm pPXF} + F_{\mathrm{Na\,\textsc{i}\,D,\,simp.}}$, and the EW is computed as $\mathrm{EW}
= \int_{\lambda_{\min}}^{\lambda_{\max}}
\left[\,1 - F_{\rm MCS}(\lambda)\,\right] \mathrm{d}\lambda$. We adopt the simple model for this step because it allows EW to take both positive (absorption) and negative (apparent emission) values under noise fluctuations, enabling a well-defined assessment of detection significance. We define detections based on the EW distribution: targets with EW $>0$ at $>2\sigma$ (i.e., 2.5th--97.5th percentile range) are classified as detections, those with EW $>0$ at $1$--$2\sigma$ (i.e., 16th--84th percentile range) as tentative detections, and those consistent with EW $\leq 0$ at $1\sigma$ as non-detections. Compared to a detection criterion based solely on per-pixel spectral S/N, this approach evaluates the significance of the integrated line strength and therefore accounts for both the line width and profile shape, providing a more robust definition of faint absorption features.

This method yields 10/23 $ 2\sigma$ detections and 3/23 $ 1\sigma$ tentative detections with excess \nai~D absorption.

Finally, to obtain accurate constraints for the absorption profile uncertainties and degeneracies, we fitted the 13 excess \nai~D absorption (tentatively) detected spectra with the partial covering model (Eq. \ref{eq: physical model}, \ref{eq: tau}, \ref{eq: full model}), using a 40000-step Markov Chain Monte Carlo (MCMC) ensemble sampler powered by \texttt{EMCEE} \citep{foreman-mackeyEmceeMCMCHammer2013}. The walkers are initialized in small regions centered on the best-fitting values obtained from the least-squares fit. We additionally fitted DD-229, including the possible $F_{\mathrm{Na\,\textsc{i}\,D,\,emi.}}$ from the P-Cygni profile; it is the only target in the \DD\ sample to host a possible emitting \nai~D component with average flux density larger than $1\sigma$ and strength comparable to the absorption feature (see Fig. \ref{fig:spec} lower right panels). When fitting with the emission component, the EW estimation increases, the absorption \voff\ estimation decreases, and the velocity dispersion increases, resulting in a $\sim0.4$ dex higher mass outflow rate and similar outflow velocity (see Table \ref{tab:outflow_properties}). Those differences result from fitting the emission feature infilling the absorption, consistent with \nai~D P-Cygni fitting attempts on spatially resolved local galaxies \citep[e.g.,][]{baronNotWindyAll2024}. We thus note the importance of including the P-Cygni emission component in deeper, higher-resolution data. The statistical tests yield $\Delta$BIC = 4.28 (defined as $\rm BIC_{abs.} - BIC_{emi.+abs.}$). According to commonly used interpretive guidelines  \citep{burnhamMultimodelInferenceUnderstanding2004}, values of $2<\Delta\rm BIC<10$ indicate moderate evidence in favor of the emission+absorption model, though the support is not strong. Furthermore, none of the outflow parameters derived with/without the emission component are significantly different ($>1\sigma$) in our case. For consistency, we use the absorption-only fit results for DD-229 in subsequent analyses. Table \ref{tab:outflow_properties} presents the best-fitting line parameters as the medians of the \texttt{EMCEE} posterior distributions, with the uncertainties corresponding to the 16th–84th percentile range. Fig. \ref{fig:spec} shows the \nai~D region (5820\AA-5960\AA) of the \nai~D-detected DeepDive spectra with the best-fitting \texttt{pPXF} continuum, and the continuum normalized spectra with the best-fitting partial covering model (Eq. \ref{eq: physical model}).

\subsection{Deriving the mass outflow rate}
\label{sec:mass_rate_derive}

We adopted the same mass outflow rate estimation as in \citet{daviesJWSTRevealsWidespread2024} (hereafter DBP24), based on the time-averaged shell model presented by \citet{rupkeOutflowsInfraredLuminousStarbursts2005} and subsequently updated by \citet{baronMultiphaseOutflowsPoststarburst2022}:

\begin{equation}
\label{eq: dotMout}
    \begin{aligned}
    \dot M_{\rm out}\,(\mathrm{M_{\odot}\,yr^{-1}}) = &11.45 \left(C_{\Omega}\frac{C_f}{0.4}\right)
    \left(\frac{N(\hi)}{10^{21}\,\mathrm{cm^{-2}}}\right) \\
    &\times \left(\frac{r_{\rm out}}{1\,\mathrm{kpc}}\right)
    \left(\frac{v_{\rm out}}{200\,\mathrm{km\,s^{-1}}}\right)
    \end{aligned}
\end{equation}

Here $C_{\Omega}$ is the large-scale covering factor set by the wind opening angle, $N(\hi)$ is the hydrogen column density, $r_{\rm out}$ is the outflow radius, and $v_{\rm out}$ is the outflow velocity \citepalias[we define $v_{\rm out} = |\Delta v| + 2\sigma$, following][]{daviesJWSTRevealsWidespread2024}. The small-scale covering fraction $C_f$ is obtained from the partial-covering model fit (Eq. \ref{eq: physical model}). We assume $C_{\Omega}=0.5$, i.e., the outflow covers half of the solid sphere, consistent with results from local infrared galaxies \citep[e.g.,][]{rupkeOutflowsInfraredLuminousStarbursts2005}. \citetalias{daviesJWSTRevealsWidespread2024} argued for a factor-of-two systematic uncertainty in $C_f$ (i.e.\ $0.25$--$1$) based on their $\sim$25\% neutral outflow detection rate in $z\sim2$ massive galaxies, consistent with our detection of 30\% outflow detection in the \DD\ sample (see Section \ref{sec: overview}). We adopt the same uncertainty range, using the effective radius (see Section \ref{sec: radius}) as the outflow radius. Local spatially resolved studies suggest that the \nai~D outflows typically extend from $\sim1$ kpc up to 15 kpc \citep[e.g.,][]{rupkeMultiphaseStructurePower2013, rupkeSpatiallyExtendedNa2015, rupkeQuasarmodeFeedbackNearby2017, pernaMultiphaseOutflowsMkn2019, baronMultiphaseOutflowsPoststarburst2020, roberts-borsaniOutflowsStarformingGalaxies2020, averyCoolOutflowsMaNGA2022, baronMultiphaseOutflowsPoststarburst2022, rubinKinematicsColdMetalenriched2022}. Two spatially resolved \nai~D traced outflows at Cosmic Noon have sizes $\leq1$ kpc \citep{cresciBubblesOutflowsNovel2023, veilleuxFirstResultsJWST2023} and 2.7 kpc \citep{deugenioFastrotatorPoststarburstGalaxy2024}, respectively. Given the relatively compact nature of the \DD\ targets (see Table \ref{tab:outflow_properties}), using \reff\ as the outflow radius is a conservative choice and can potentially underestimate the mass outflow rates by up to 1 dex.

We estimated the hydrogen column density as:
\begin{equation}
    N(\hi) = \frac{N(\nai)}{(1-y)10^{-(a+b)}}
\end{equation}
\noindent
where $a$ is the sodium abundance term, $b$ is the dust depletion factor, and $N(\nai)$ is the \nai~column density. Following \citet{shihComplexStructureMultiphase2010}, we assume $a = \log [N_{\rm Na}/N_{\rm H}] = -5.69$ and $b = \log [N_{\rm Na}/N_{\rm H, total}] - \log [N_{\rm Na}/N_{\rm H, gas}] = -0.95$. Following most previous works, we assumed a Milky-Way-like fraction of 0.1 for the sodium neutral fraction $(1-y)$ \citep{shihComplexStructureMultiphase2010, rupkeSpatiallyExtendedNa2015, rupkeQuasarmodeFeedbackNearby2017, daviesJWSTRevealsWidespread2024}. However, \citet{baronMultiphaseOutflowsPoststarburst2020} measured a 0.05 sodium neutral fraction in a local AGN-driven outflow; it may be that the neutral fraction is underestimated in more extreme cases. Consequently, the hydrogen column densities, and thus also the mass outflow rates, would increase by a factor of two. Similarly, \citet{morettiEmpiricalCalibrationNa2026} shows that with Milky-Way-like assumptions, one can overestimate the \hi~column density by 30\% for $z>2$ galaxies.

We estimated the \nai~column density with the optical depth at the center of the \nai~D1 line, $\tau_{\rm 0,\,D1}$, via:

\begin{equation}
\begin{aligned}
    \mathrm{N(\nai)} \, (\mathrm{cm^{-2})} = &10^{13} \left(\frac{\tau_{\rm 0,\,D1}}{0.7580}\right)\left(\frac{0.4164}{f_{\rm lu}}\right)\\
                         &\times \left(\frac{1215\AA}{\lambda_{\rm lu}}\right) \left(\frac{b}{10\rm \,km\,s^{-1}}\right)
\end{aligned}
\end{equation}
\noindent
where $f_{\rm lu}=0.32$ and $\lambda_{\rm lu}=5897.6\AA$ are the oscillator strength and rest-frame wavelength of the transition, respectively \citep[see e.g.,][]{drainePhysicsInterstellarIntergalactic2011}. $b$ is the Doppler parameter, related to the velocity dispersion by $b=\sqrt{2}\sigma$. 

We derived the mass outflow rates from the full Monte Carlo posterior probability distributions of $\Delta v$, $\sigma$, $C_f$, and $\tau_{0,\,\mathrm{D1}}$. The $C_f$ and $\tau_{0,\,\mathrm{D1}}$ parameters are degenerate because the \nai~D lines are blended in our spectra. Nevertheless, since the mass outflow rate depends on the product of these two parameters (Eq.~\ref{eq: dotMout}), the posterior distributions of the derived mass outflow rates remain well constrained \citepalias[as in][]{daviesJWSTRevealsWidespread2024}.

\begin{figure}
    \centering
    \includegraphics[width=0.9\linewidth]{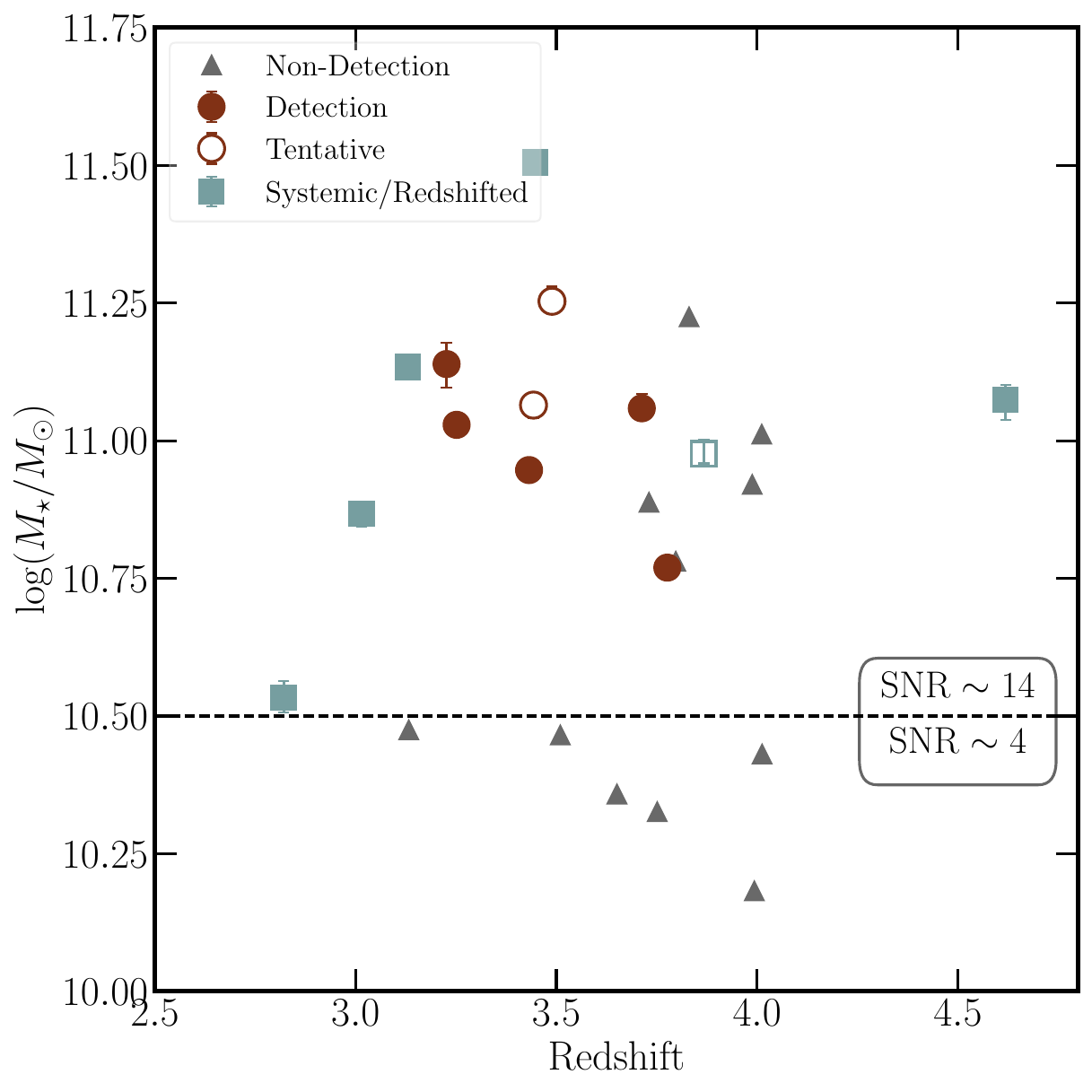}
    \caption{\DD\ QGs plotted on the redshift--stellar mass plane. The \nai~D blueshift (tentatively) detected targets are in (hollow) red circles, and the systemic targets are in cyan squares. The \nai~D excess absorption non-detections are plotted as gray triangles. No targets are detected below a stellar mass threshold of $10^{10.5}M_\odot$.
    The median S/N ratios for galaxies above and below this threshold are indicated.}
    \label{fig:z_vs_M}
\end{figure}

\begin{table*}
\caption{Outflow properties of the \DD\, targets}
\label{tab:outflow_properties}
\centering
\begingroup
\setlength{\tabcolsep}{3pt}
\begin{tabular}{@{}ccccccccccc@{}}
\hline\\[0.5pt]
(1) DD-ID & (2) RA & (3) Dec. & (4) $z_{\mathrm{spec}}$ 
& (5) $\rm SFR_{100\,Myr}$ & (6) $r_{\rm eff}$ 
& (7) EW & (8) \voff & (9) $\sigma_{\rm \nai}$ & (10) $v_{\mathrm{out}}$ 
& (11) $\log(\dot{M}_{\mathrm{out}})$\\[3pt]

& deg & deg &  & $[\mathrm{M_\odot\,yr^{-1}}]$ & [kpc] 
& [\AA] & [km s$^{-1}$] & [km s$^{-1}$] & [km s$^{-1}$] 
& $[\mathrm{M_\odot\,yr^{-1}}]$\\[3pt]
\hline\\[1pt]

\multicolumn{11}{c}{\textit{Blueshifted}} \\[6pt]

115 & 149.41959 & 2.00753 & 3.78 
    & $16.18_{-8.56}^{+11.50}$ & $0.72 \pm 0.02$
    & $1.6_{-0.5}^{+0.5}$ 
    & $-324_{-112}^{+90}$ & $78_{-54}^{+73}$ & $480_{-141}^{+180}$ 
    & $1.24_{-0.37}^{+0.39}$ \\[6pt]

134\tablefootmark{a} & 150.06147 & 2.37871 & 3.71 
    & $0.01_{-0.01}^{+0.16}$ & $0.69 \pm 0.05$
    & $6.1_{-0.4}^{+0.4}$ 
    & $-113_{-17}^{+18}$ & $58_{-13}^{+16}$ & $230_{-30}^{+36}$ 
    & $1.65_{-0.25}^{+0.23}$ \\[6pt]

170 & 36.73354 & -4.53658 & 3.49 
    & $0.57_{-0.56}^{+2.03}$ & $1.63 \pm 0.02$
    & $1.0_{-0.5}^{+0.5}$ 
    & $-233_{-140}^{+169}$ & $74_{-58}^{+99}$ & $381_{-138}^{+212}$ 
    & $1.33_{-0.49}^{+0.44}$ \\[6pt]

185\tablefootmark{b} 
    & 214.86605 & 52.88409 & 3.44 
    & $2.83_{-1.35}^{+2.03}$ & $1.47 \pm 0.02$
    & $2.5_{-1.0}^{+1.0}$ 
    & $-453_{-192}^{+169}$ & $138_{-95}^{+75}$ & $730_{-248}^{+211}$ 
    & $2.00_{-0.47}^{+0.43}$ \\[6pt]

196\tablefootmark{b} 
    & 214.86605 & 52.88426 & 3.43 
    & $0.01_{-0.01}^{+0.15}$ & $0.53 \pm 0.03$
    & $1.7_{-0.6}^{+0.6}$ 
    & $-673_{-174}^{+304}$ & $192_{-129}^{+67}$ & $1057_{-278}^{+110}$ 
    & $1.65_{-0.45}^{+0.39}$ \\[6pt]

236 & 34.42766 & -5.15242 & 3.23 
    & $25.29_{-10.30}^{+18.09}$ & $3.96 \pm 0.16$
    & $7.2_{-0.7}^{+0.7}$ 
    & $-166_{-35}^{+35}$ & $108_{-23}^{+26}$ & $382_{-53}^{+58}$ 
    & $2.68_{-0.28}^{+0.27}$ \\[8pt]

\multicolumn{11}{c}{\textit{P-Cygni}} \\[6pt]

\multirow{2}{*}{229\tablefootmark{c}} 
    & \multirow{2}{*}{214.89561} & \multirow{2}{*}{52.85650} & \multirow{2}{*}{3.25} 
    & \multirow{2}{*}{$0.74_{-0.32}^{+0.50}$} 
    & \multirow{2}{*}{$0.79 \pm 0.03$}
    & $1.3_{-0.3}^{+0.4}$ {\scriptsize (abs.)} 
    & $-539_{-88}^{+88}$ & $105_{-74}^{+61}$ & $750_{-151}^{+145}$ 
    & $1.45_{-0.37}^{+0.36}$ \\[2pt]

& & & & 
    & 
    & $1.8_{-0.6}^{+0.8}$ {\scriptsize (emi.+abs.)} 
    & $-466_{-104}^{+132}$ & $155_{-65}^{+63}$ & $754_{-135}^{+136}$ 
    & $1.78_{-0.46}^{+0.39}$ \\[8pt]

\multicolumn{11}{c}{\textit{Systemic}\tablefootmark{d}} \\[6pt]

96  & 34.75628 & -5.30809 & 3.87 
    & $0.01_{-0.01}^{+0.09}$ & $0.54 \pm 0.02$
    & $1.1_{-0.7}^{+0.7}$ 
    & $-71_{-390}^{+119}$ & $64_{-49}^{+134}$ & $198_{-104}^{+597}$ 
    & $0.67_{-0.57}^{+0.68}$ \\[6pt]

179 & 34.38699 & -5.48269 & 3.45 
    & $0.12_{-0.12}^{+0.83}$ & $1.96 \pm 0.03$
    & $1.5_{-0.4}^{+0.5}$ 
    & $-7_{-132}^{+118}$ & $88_{-63}^{+88}$ & $183_{-39}^{+279}$ 
    & $1.40_{-0.45}^{+0.47}$ \\[6pt]

257\tablefootmark{e} & 34.29108 & -5.03812 & 3.13 
    & $28.80_{-4.07}^{+4.19}$ & $2.12 \pm 0.04$
    & $4.8_{-0.4}^{+0.4}$ 
    & $-27_{-41}^{+40}$ & $175_{-33}^{+41}$ & $377_{-58}^{+98}$ 
    & $1.69_{-0.29}^{+0.30}$ \\[6pt]

305 & 150.12752 & 2.35977 & 2.82 
    & $0.06_{-0.06}^{+0.27}$ & $0.51 \pm 0.01$
    & $4.1_{-0.6}^{+0.6}$ 
    & $8_{-34}^{+35}$ & $43_{-14}^{+24}$ & $94_{-15}^{+75}$ 
    & $0.98_{-0.27}^{+0.26}$ \\[8pt]

\multicolumn{11}{c}{\textit{Redshfited}\tablefootmark{d}} \\[6pt]

53\tablefootmark{f} & 34.39968 & -5.13635 & 4.62 
   & $0.10_{-0.09}^{+0.68}$ & $0.61 \pm 0.08$
   & $3.0_{-0.7}^{+0.5}$ 
   & $104_{-37}^{+37}$ & $71_{-42}^{+66}$ & $245_{-90}^{+140}$ 
   & $1.18_{-0.31}^{+0.32}$ \\[6pt]

270 & 34.78588 & -5.35732 & 3.02 
    & $0.77_{-0.35}^{+0.53}$ & $1.15 \pm 0.04$
    & $1.9_{-0.3}^{+0.3}$ 
    & $86_{-58}^{+56}$ & $82_{-46}^{+49}$ & $250_{-94}^{+113}$ 
    & $1.20_{-0.32}^{+0.33}$ \\[6pt]

\hline
\end{tabular}
\endgroup

\tablefoot{(1) \DD\ IDs;
(2) RA; (3) Dec.; (4) Spectroscopic redshift; 
(5) $\rm SFR_{100\,Myr}$ \citep{hamadoucheDeepDiveTracingEarly2026};
(6) Effective semi-major axis; 
(7) \nai~D rest-frame equivalent width; 
(8) Line-center velocity offset; 
(9) Velocity dispersion of the excess \nai~D absorption; 
(10) Outflow velocity, defined as $|\Delta v|+2\sigma$ following \citetalias{daviesJWSTRevealsWidespread2024}; 
(11) Mass outflow rate.\\[3pt]
\tablefoottext{a}{DD-134 is a member of the QG-SFG merging pair ``Jekyll and Hyde'' \citep{schreiberJekyllHydeQuiescence2018}.}
\tablefoottext{b}{DD-185 and DD-196 are a QG-QG merging pair \citep{itoMergingPairMassive2025}.}
\tablefoottext{c}{DD-229 has two measurements: absorption-only and emission+absorption.}
\tablefoottext{d}{We derive ``outflow'' properties for the systemic/redshifted excess \nai~D absorption detected targets and include them in further analysis for completeness.}
\tablefoottext{e}{
\reff based on the NIRCam F356W image.}
\tablefoottext{f}{\citet{taylorJWSTEXCELSSurvey2026} also reported DD-53 (EXCELS-117560) with redshifted \nai~D of \voff$=100^{+70}_{-60}\rm\,\,km\,s^{-1}$.}
}

\end{table*}

\section{Outflow properties}
\label{sec: results}

\subsection{Overview}
\label{sec: overview}

Table \ref{tab:outflow_properties} summarizes the outflow properties for all targets, and Fig. \ref{fig:z_vs_M} gives an overview of the \nai~D detection on the redshift-stellar mass plane. We identify interstellar \nai~D absorption in 13 out of 23 \DD\ QGs (57\%), of which ten are detected at the $2\sigma$ level, and three are $1\sigma$ tentative detections. Four galaxies show excess absorption consistent with the systemic velocity (i.e., $1\sigma_{\Delta v}>0$ \kms), two galaxies show potential redshifted excess absorption with positive low velocity offset ($\Delta v\sim90\,\,\rm km\,s^{-1}$), while the remaining 7/23 or 30\% (tentatively) detected sources exhibit blueshifted excess \nai~D absorption with $\Delta v \lesssim -150$ \kms, indicating outflowing neutral gas. Our detection fraction of the (either blueshifted or not) excess \nai~D absorption is slightly higher than that of the $z\sim2$ massive sample in \citetalias{daviesJWSTRevealsWidespread2024}, and also higher than that of a recent $z\sim3$ post-starburst galaxy (PSB) sample \citep{taylorJWSTEXCELSSurvey2026}, refer to Section \ref{sec: compare} for details. As shown in Fig. \ref{fig:z_vs_M}, the \nai~D excess absorption (tentatively) detected targets (red circles and cyan squares) are all massive ($\log(M_\star/M_\odot)>10.5$). The detection of outflows on the massive end is consistent with lower redshift studies \citepalias{sunEvolutionGasFlows2024, daviesJWSTRevealsWidespread2024}, and also with the recent statistical study across $1<z<10$ \citep{lyuFirstStatisticalDetection2026}. However, we note that this may reflect a brightness bias: massive galaxies tend to be brighter and easier to detect. The lower mass ($\log(M_\star/M_\odot)<10.5$) \DD\ targets have a median spectral S/N of 3.7, while the more massive ones have a median spectral S/N of 14. 

The two targets with the highest \voff\ (DD-185 and DD-196) are reported by \citet{itoMergingPairMassive2025} as a merging pair, suggesting that their high \voff\ may be aided by an ongoing merger. The neutral outflow in DD-134, also in a close physical pair with a dusty star-forming companion (\textit{Jekyll \& Hyde} in \citealt{schreiberJekyllHydeQuiescence2018}), might also be partially associated with the other member of the pair based on spatially resolved observations \citep{perez-gonzalezAcceleratedQuenchingChemical2025}. If we strictly exclude these possible merger-related targets, we are left with 4/23 (17\%) targets with blueshifted outflow signs.

Given the $\sim100$ \kms\ resolving power of \nirspec-G235M and the expected dependence of \voff\ on outflow geometry, the systemic ($1\sigma_{\Delta v}>0$ \kms) and low-$\Delta v$ redshifted excess \nai~D absorption detected targets may still host some outflowing sodium. For completeness, we include the systemic and redshifted detections in subsequent analyses, including the derivation of outflow velocities and mass outflow rates, and label them as cyan squares. Moreover, one target, DD-229, shows a possible P-Cygni profile (i.e., blueshifted absorption paired with systemic or redshifted emission features). \citet{prochaskaSimpleModelsMetalline2011} suggest that for galactic outflows traced by resonance transitions (e.g., \mgii, \nai, \feii), depending on the geometry, a P-Cygni profile can be observed and often relates to the most substantial outflow. In the local universe, there are reported cases in the literature of \nai~D P-Cygni profiles in spatially resolved systems \citep[e.g.,][]{rupkeSpatiallyExtendedNa2015, pernaMultiphaseOutflowsMkn2019, baronMultiphaseOutflowsPoststarburst2020, baronNotWindyAll2024}. Moreover, in a large SDSS sample, \citetalias{sunEvolutionGasFlows2024} find that $\sim7\%$ of the flow-detectable PSBs have P-Cygni profiles. Altogether, this suggests that P-Cygni profiles associated with outflows are not uncommon. However, beyond the local universe, to our knowledge, there have been no reports of convincing observational evidence for P-Cygni profiles. In our case, the possible P-Cygni host DD-229 also exhibits a broad \halpha\ component ($\sigma_{\rm H\alpha -broad}\sim4900\,\,\rm km\,s^{-1}$) and  $\log(\oiii/H\beta)=0.86^{+0.38}_{-2.67}$, it is also X-ray confirmed (see Section \ref{sec: AGN}), providing compelling evidence for an AGN. DD-229's outflow velocity offset, measured with/without the emission component, is the highest among all targets except the merging pair DD-185 and DD-196. These suggest a strong AGN-driven outflow in DD-229.

\begin{figure*}[h]
    \centering
    \includegraphics[width=\linewidth]{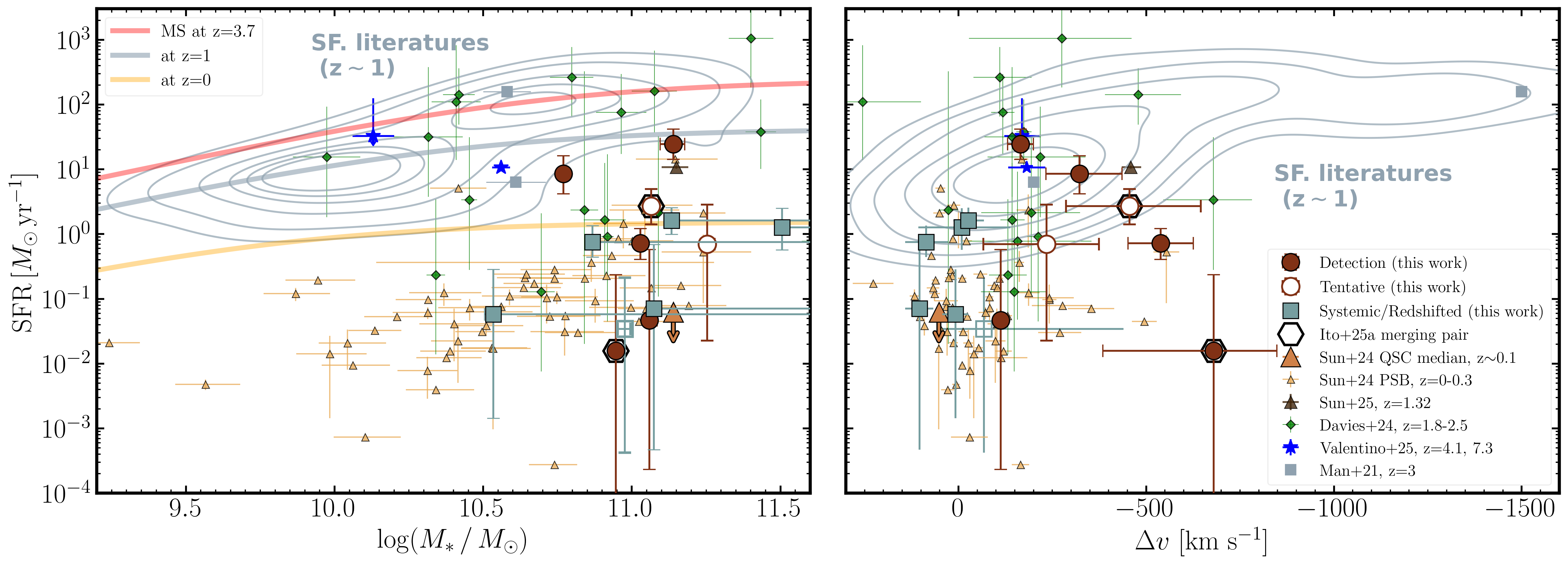}
    \caption{$\mathrm{SFR_{100\,Myr}}$ as a function of stellar mass and \nai\,D (or \mgii\, for the gray markers) velocity offset. The red hollow/filled circles are the blueshifted \nai\,D (tentatively) detected DD targets \citep[black hexagons frame the merging pair in][]{itoMergingPairMassive2025}, and the cyan squares are the systemic/redshifted \nai\,D targets. The big orange and small yellow triangles represent the local quiescent (stacked with SFR upper limits) and post-starburst samples in \citetalias{sunEvolutionGasFlows2024}, respectively. The brown big triangle represents the recently quenched source with extreme outflow in \citet{sunExtremeNeutralOutflow2026}. The green diamonds represent the excess \nai~D sample (including the blueshifted/systemic/redshifted targets) from \citetalias{daviesJWSTRevealsWidespread2024}. The blue stars are the two recently quenched high-z galaxies reported by \citet{valentinoGasOutflowsTwo2025}. The gray contours represent the homogenized literature compilation of \mgii\ outflows across redshifts \citep[][and references therein]{davisExtendingDynamicRange2023}.
    The location of two $z\sim3$ recently quenched sources in \citet{manExquisitelyDeepView2021} is marked by gray squares. On the left panel, we plot the star-forming main sequences \citep[see][equation 14]{popessoMainSequenceStarforming2023} at $z=3.7$ (median \DD\ redshift), $z=1$, and $z=0$ as light red/gray/orange lines, respectively.}
    \label{fig:mass_vel_sfr}
\end{figure*}

\begin{figure*}
    \centering
    \includegraphics[width=0.9\linewidth]{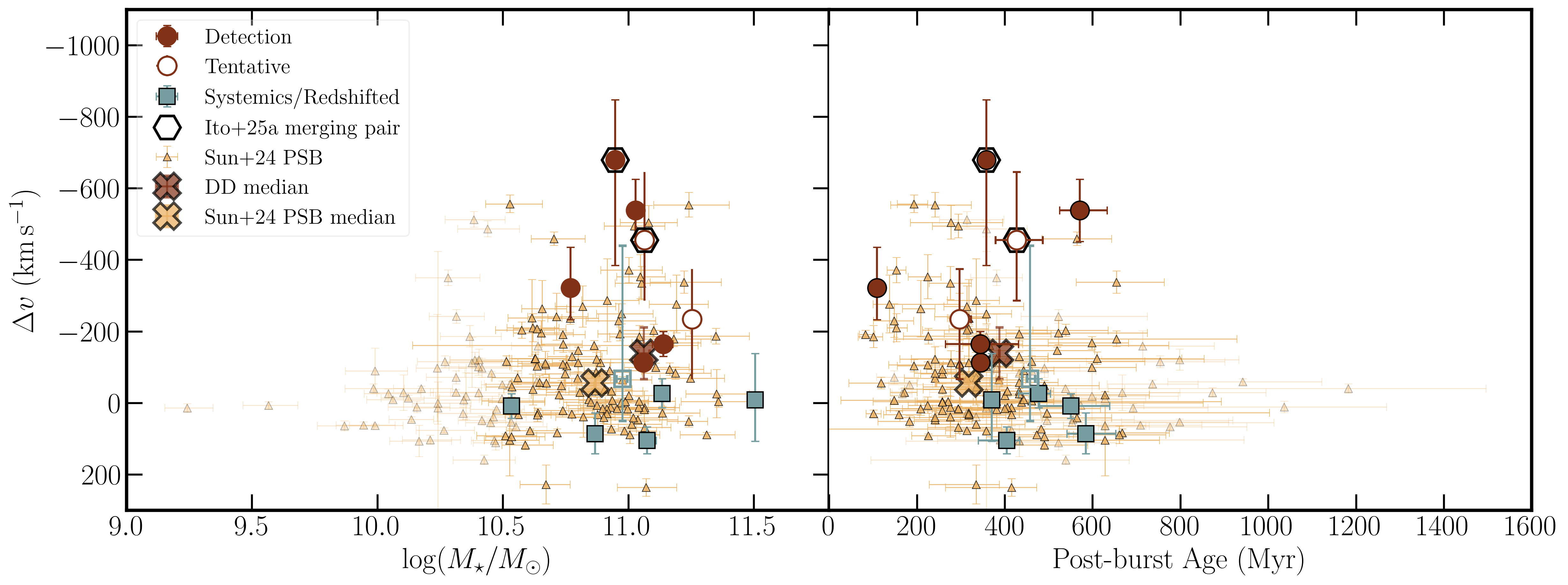}
    \caption{The \nai~D velocity offset \voff\ as a function of stellar mass (\textit{Left}), and post-burst age (time passed since 90\% of $M_\star$ formed, \textit{Right}). The blueshifted \nai~D (tentatively) detected \DD\ targets are plotted as hollow/red circles, and the systemic targets as cyan squares. As a comparison, \citetalias{sunEvolutionGasFlows2024} local PSBs are plotted as yellow triangles. The median stack of the \DD\ targets (including the merger and the systemic targets) and of the \citetalias{sunEvolutionGasFlows2024} PSBs (only those in the same mass and age bin as \DD\ targets) are marked with colored crosses. The \DD~targets show similar \voff~as their local counterparts, and there is no clear trend in either of the figures, given the limited sample size.}
    \label{fig:age_voff}
\end{figure*}

\subsection{Comparison with the literature}
\label{sec: compare}

Comparing with previous work, \citetalias{sunEvolutionGasFlows2024} reported that 32\% of 516 local PSBs host excess sodium, and 20\% show outflowing features; \citetalias{daviesJWSTRevealsWidespread2024} identified excess \nai~D absorption in 27\% of 113 galaxies at $z\sim2$ (including both star-forming and quiescent systems), and $\sim13$\% hosts outflows, with almost all of the \nai~D detections on the massive end ($M_\star>10^{10}\,\mathrm{M_\odot}$). More recently, \citet{taylorJWSTEXCELSSurvey2026} find potential \nai~D outflows in 3/13 ($\sim23$\%) $z\sim3$ PSBs and potential inflows in 2/13; we confirm their redshifted \nai~D case EXCLES-117560 (DD-53, see Table \ref{tab:outflow_properties}). The left panel of Fig. \ref{fig:mass_vel_sfr} demonstrates how the SFR and $M_\star$ of the \DD\ targets compare to the literature samples. The \DD\ targets lie in the low-SFR and high-mass region, deviating from the $z=3.7$ star-forming main sequence shown as the red curve. The \DD\ sample is comparable with a stack of local massive QGs and individual PSB detections \citepalias[][]{sunEvolutionGasFlows2024}. The \citetalias{daviesJWSTRevealsWidespread2024} targets (green diamonds) have higher SFRs, but comparable masses with the \DD\ targets and the local PSBs/QGs. In contrast to the literature samples, the \DD\ sample has a higher detection fraction: 57\% show excess \nai~D absorption and 30\% display blueshifted outflow signatures. Although this slightly enhanced rate may reflect bias given the higher stellar masses of the \DD\ galaxies, it nevertheless indicates that neutral outflows are common among massive quiescent systems even at $z>3$.\\

In the right panel of Fig. \ref{fig:mass_vel_sfr}, we compare the \DD\ and literature samples on the \voff--SFR plane. Three \DD\ targets show \voff$<-400$ \kms, appear to be faster than most of their literature counterparts. However, we note that two of the high \voff\ targets are likely merging \citep{itoMergingPairMassive2025}, so their blueshifted excess \nai~D absorption might be affected by gravitational interactions. The other high-\voff\ target, DD-229, is a broad-line AGN with a P-Cygni \nai~D profile. The rest of the \DD\ targets appear to fall in the same region of distribution with the local PSBs and $z\sim2$ counterparts, with no obvious correlation between \voff\ and SFR (Spearman's correlation coefficient $\rho=-0.6$). The $z\sim3.5$ \DD\ sample is, in general, not so different from the lower-redshift dwellers.

Fig. \ref{fig:age_voff} compares the \DD\ sample with the \citetalias{sunEvolutionGasFlows2024} local PSBs on the \voff--$M_\star$ and the \voff--post-burst-age planes. Following \citet{frenchClockingEvolutionPoststarburst2018}, the post-burst age here is defined as the time elapsed since $90\%$ of the mass formed until observation. We recovered the \DD\ post-burst ages from the SFH modeled by \texttt{Bagpipes} \citep{hamadoucheDeepDiveTracingEarly2026}. The colored crosses in Fig. \ref{fig:age_voff} give the median value of the \DD\ targets (including the merging pair and the systemic/redshifted targets, to keep consistent with the \citetalias{sunEvolutionGasFlows2024} sample) and of the \citetalias{sunEvolutionGasFlows2024} PSBs (matched to the same mass and age bins as the \DD\ targets). The median velocity offset for the \DD\ targets is slightly higher but very similar ($<1\sigma$) to the local PSBs in \citetalias{sunEvolutionGasFlows2024} in the same mass and age bins. Both samples have a median \voff$\sim100$ \kms, suggesting that the outflow is neither more drastic nor weaker at higher redshifts than at local --- if the outflow plays a role in the quenching mechanism, it is unlikely to be redshift-dependent. On the other hand, if taking into account the higher \nai~D excess absorption detection rate (including blueshifted, redshifted, and systemic) for the \DD\ sample (57\%) compared with the \citetalias{sunEvolutionGasFlows2024} PSBs ($\sim$32\%), it is still possible that the outflow is more common in PSBs at $z>3$. However, we note that this higher detection rate for the \DD\ targets may be a result of incomplete sampling, as even when limiting to the same mass and age bins, the \citetalias{sunEvolutionGasFlows2024} sample is still $\sim$0.5 dex lower in stellar mass on average relative to the \DD\ sample. Thus, the higher \DD\ \nai~detection rate might be biased by our higher masses, as many studies, including Fig. \ref{fig:z_vs_M}, have shown a correlation between outflow and host galaxy stellar mass \citep[e.g.,][]{sunEvolutionGasFlows2024, daviesJWSTRevealsWidespread2024, lyuFirstStatisticalDetection2026}.

\begin{figure*}[t]
    \centering
    \includegraphics[width=0.45\linewidth]{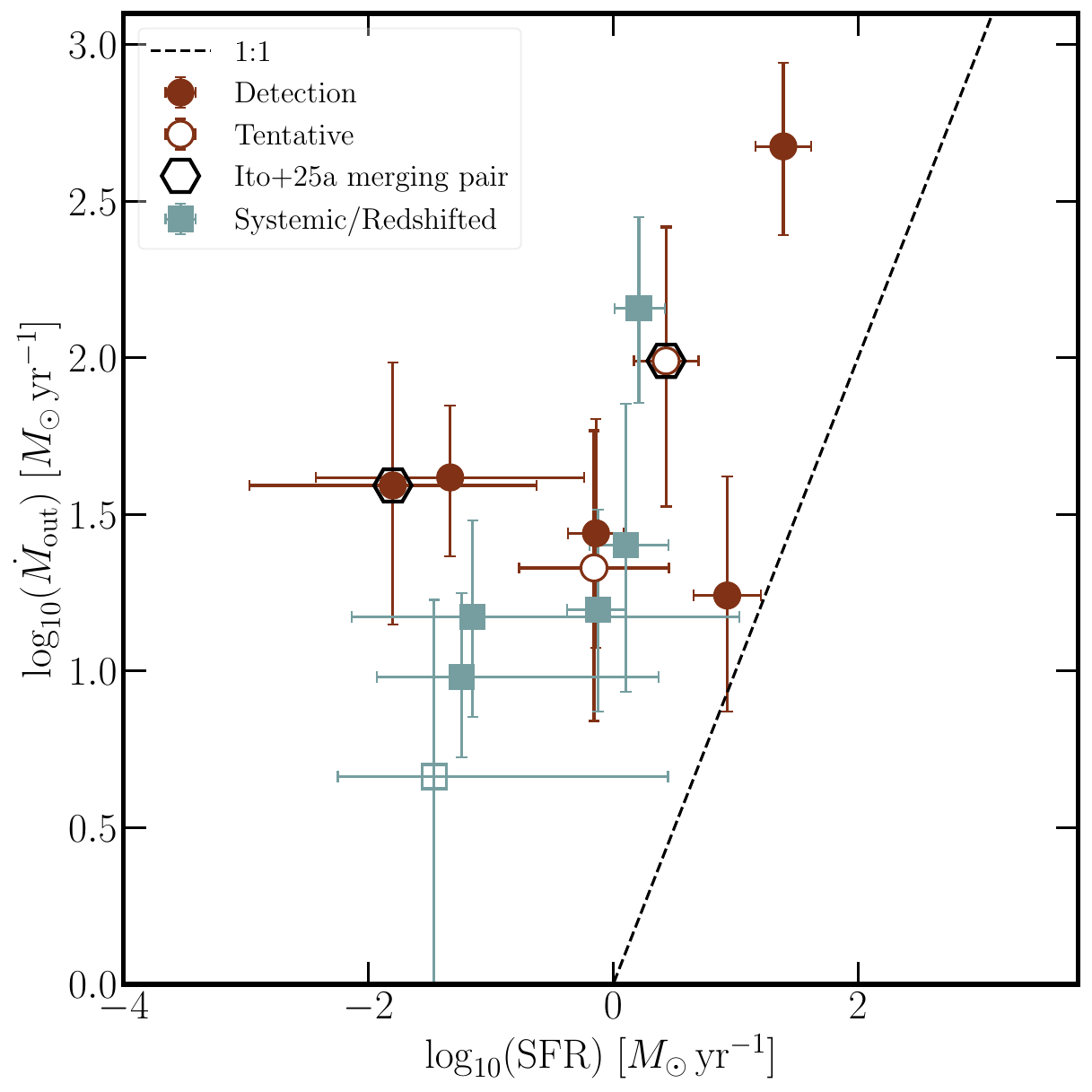}
    \hspace{0.5cm}
    \includegraphics[width=0.45\linewidth]{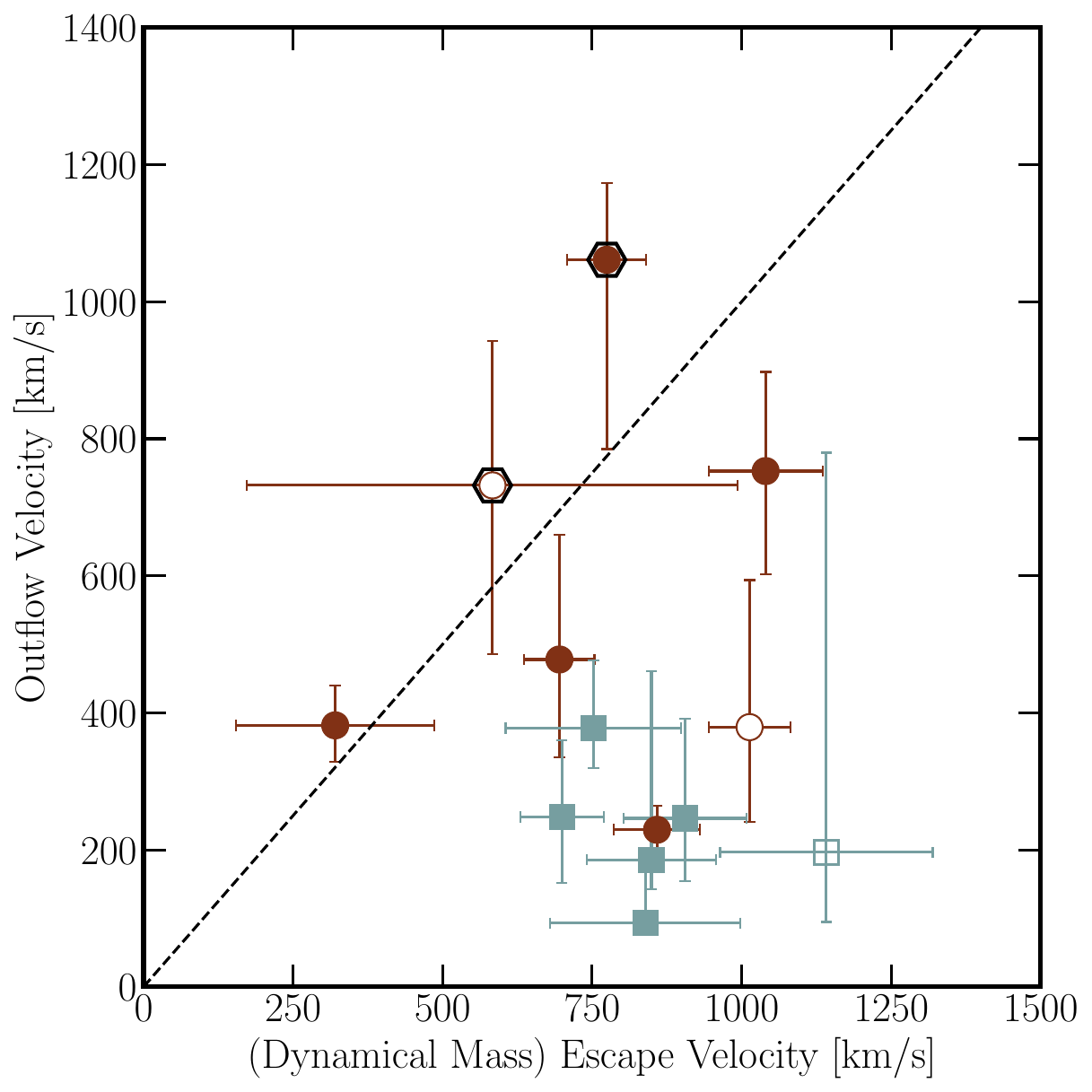}
    \caption{\textit{Left:} The $\rm SFR_{100Myr}$ vs. the mass outflow rate. All the \DD\ targets show higher mass outflow rates than the ongoing SFR, implying the neutral gas outflow is able to suppress the ongoing star formation. \textit{Right:} The dynamical mass escape velocity at \reff\ vs. the outflow velocity. Most \DD\ targets show an outflow velocity smaller than the escape velocity, meaning that the blueshifted neutral gas is not likely to escape the galaxy. Three targets show $v_{\rm out}> v_{\rm esc}$, two of which are the \citet{itoMergingPairMassive2025} merging pair, and the other one is the X-ray source DD-236 with an extreme mass outflow rate (Section \ref{sec: mass outflow rate}).}
    \label{fig:vesp_vout&sfr_mout}
\end{figure*}

\subsection{Mass outflow rates}
\label{sec: mass outflow rate}

The left panel of Fig. \ref{fig:vesp_vout&sfr_mout} gives a comparison between the mass outflow rate and the ongoing SFR (averaged over the last 100 Myr). The mass outflow rates seem to increase with higher SFRs, but show only a weak correlation (Spearman's correlation coefficient of $\rho=0.46$, p-value=0.11). This possible \DD\ SFR--$\dot{M}_{\rm out}$ weak correlation is broadly consistent with lower-redshift studies \citep[e.g.,][]{rupkeOutflowsInfraredLuminousStarbursts2005, fioreAGNWindScaling2017, sunEvolutionGasFlows2024}. For all the \DD\ targets, their mass outflow rates are orders of magnitude greater than the ongoing SFRs. \citet{belliStarFormationShut2024} reported similar findings for a $z=2.45$ PSB, and suggest that the mass outflow rate is thus capable of quenching or at least suppressing the star formation. Other single-object or sample studies at $z\sim2-4$ find even stronger enhancements in the mass outflow rates of QGs \citep[e.g.,][]{daviesJWSTRevealsWidespread2024, wuEjectiveFeedbackQuenching2025, taylorJWSTEXCELSSurvey2026}. However, to permanently quench a galaxy, the mass budget carried by the neutral outflows must be removed from the host galaxy; i.e., the outflows must be fast enough to escape. Otherwise, the high mass outflow rates can only suggest active fountain-like mechanisms. We discuss the outflow's escape in the next section.

One target, DD-236, has a remarkably high mass outflow rate (\dMout$=2.68^{+0.27}_{-0.28}$) -- the highest reported beyond the local universe to date. DD-236 is not particularly bright: it has an F200W AB magnitude of 23.1, lower than the median \DD\ F200W magnitude of 22.5, resulting in a S/N of 10 in the G235M spectrum. Nevertheless, DD-236 shows a very deep excess \nai~D absorption feature (Fig. \ref{fig:spec}). The NIRCam image shows no close companions near DD-236 that could be potential merger partners, ruling out such contamination to the excess \nai~D absorption feature. Interestingly, the spectrum shows significant \oiii\ emission, but no \hbeta.  While \nii\ and \halpha\ fall in a detector gap, thus precluding our ability to constrain the presence of AGN with the classical BPT diagnostics (Section \ref{sec: AGN}), the upper limit on its \oiii/\hbeta\ ratio ($\log(\oiii/H\beta)<0.77$) for the high stellar mass ($M_\star = 10^{11.14}\,\rm M_\odot$) is consistent with an AGN source using the MEx diagnostic \citep{juneauNewDiagnosticActive2011}. Moreover, DD-236 is detected in X-ray ($L_{\rm 2-10\,keV}=5\times10^{43}\,\mathrm{erg}\,\mathrm{s^{-1}}$, \citealt{kocevskiXUDSChandraLegacy2018}), further supporting the presence of nuclear activity, which might be linked to the presence of a massive outflow, as observed in the local Universe \citep[e.g.,][]{harrisonImpactSupermassiveBlack2017, veilleuxCoolOutflowsGalaxies2020, baronMultiphaseOutflowsPoststarburst2022}.

With all that said, we note that the mass outflow rate derived from unresolved excess \nai~D absorption features relies on several assumptions and is thus affected by large systematic uncertainties potentially amounting to $\sim1$ dex (see also Section \ref{sec: discuss}).

\section{Escaping or recycling}
\label{sec: escape}

To understand if these gaseous ``out''-flows are strong enough to leave their hosts, we estimate the escape velocities for the \DD\ targets using their dynamical mass ($M_{\mathrm{dyn}}$) inferred from the \texttt{pPXF} stellar velocity dispersions, following Equation 20 in \citet{cappellariSAURONProjectIV2006}. $M_{\mathrm{dyn}}$ is a good representation of the baryon-dominated mass within 2\reff. For each target, we assume a simple profile in which half of the dynamical mass is enclosed within $ \reff$, and estimate the escape velocity at \reff\ as $v_{\mathrm{esc}} = \sqrt{2\,G*0.5M_{\mathrm{dyn}}/r_{\mathrm{eff}}}$ to be consistent with the assumption adopted to compute $\dot{M}_{\rm out}$ in the previous section. We note that the \DD\ targets are compact and massive, thus having a relatively high escape velocity. The right-hand panel of Fig. \ref{fig:vesp_vout&sfr_mout} shows the comparison between outflow velocity \citepalias[defined as $v_{\rm out} = |\Delta v|+2\sigma$, following][]{daviesJWSTRevealsWidespread2024} and the escape velocity. Except for the X-ray AGN DD-236 (also with very high $\dot{M}_{\rm out}$, see Section \ref{sec: mass outflow rate}) and the mergers DD-186 and DD-196, all other outflows have velocities below the escape velocity. Considering that the $v_{\rm esc}$ estimated here is a lower limit of the total escape velocity, it is likely that most of the \DD\ outflows cannot escape their hosts and may be recycled in some way.

Using toy models implemented with \texttt{galpy} \citep{bovyGalpyPythonLibrary2015}, we estimated the dynamical ``fountain'' timescale (i.e., the time it takes from the launching radius to return to \reff) of the outflowing neutral gas under two gravitational potentials: (1) A Keplerian potential, assuming the dynamical mass is evenly distributed within 2\reff\ and purely ballistic motion; (2) A Navarro--Frenk--White (NFW) potential \citep[i.e., that of a dark matter halo][]{navarroUniversalDensityProfile1997}, with a representative virial mass of $M_{\rm vir} = 10^{13}\,M_\odot$ for $z>3$ massive quiescent galaxies \citep{behrooziUniverseMachineCorrelationGalaxy2019}, adopting the concentration--mass relation from \citet{duttonColdDarkMatter2014}.
The Keplerian potential neglects the additional gravitational contributions from the extended dark matter halo and the circumgalactic medium (CGM), as well as neglects hydrodynamic drag. Thus, it provides a lower bound on the return timescale. The NFW model is a reference to account for the expected dark matter halos hosting our massive \DD\ systems.

Following the assumptions in Section \ref{sec:mass_rate_derive} that the observed outflow lunches at $r_{\rm out}=\reff$, we find that, except for the merging pair (DD-185, DD-196) and DD-236 with $v_{\rm out} > v_{\rm esc}$, none of the other \DD\ targets hosts a strong enough outflow that can escape the hosting galaxy. The fountain outflows of other \DD\ targets have, on average, $t_{\rm return}\sim3$ Myr assuming the Keplerian potential. In the case of NFW potential, it increases the timescale to $\sim 115$ Myr. These timescales are 0.4-2 dex lower compared with the median \texttt{Bagpipes} post-burst age of $\sim404\pm28$ Myr, suggesting that these outflows are unlikely to correlate with the most recent starburst episodes, as they cycle on relatively short timescales.

If we consider the most extreme cases in the local universe where the excess \nai~D absorption is observed at 15 kpc in local quasar spectra \citep{rubinKinematicsColdMetalenriched2022}: assuming gas is launched with $v_{\rm out}$ at $r_{\rm out} = 15$ kpc, all the detected \DD\ outflows become unbound in the Keplerian potential; and all are bound in the NFW potential with a median $t_{\rm return}\sim180$ Myr. If we assume a somewhat middle point value of $r_{\rm out} = 7$ kpc (see Section \ref{sec:mass_rate_derive} for a discussion on the outflow radius range), in the Keplerian potential $\sim$half of the detected \DD\ outflows can escape, and the rest have a median return timescale of $\sim 60$ Myr; in the NFW potential all are still bound with $t_{\rm return}\sim140$ Myr.

In conclusion, depending on the assumed outflow radius and potentials, we find: (1) For the Keplerian potential, the outflows' escape capabilities and return timescales highly depends on the launching radius, $t_{\rm return}$ can vary from short timescales of $\sim3$ Myr when launching at \reff up-to infinite when launching at 15 kpc; (2) In the NFW potential, accounting for the dark matter halo of a typical massive distant quenched galaxy of $M_\star\sim10^{11}\,M_\odot$, the return timescales at different $r_{\rm out}$ do not show significant differences, yielding $t_{\rm return}\sim110-180$ Myr, and the outflow cannot escape the potential in any case. Depending on $r_{\rm out}$, the outflow is capable of escaping the inner, stellar body of the galaxy, but always remains gravitationally bound in the dark matter halo. Moreover, for both potentials if $r_{\rm out}\lesssim7$ kpc, the return timescale is relatively short with $t_{\rm return}\lesssim150$ Myr. Our findings imply that the outflow can remove gas from the inner region, possibly into the CGM if launched from a sufficiently large radius; in the case of inner-region-bound gas, it ``fountains'' back relatively quickly.

Considering the launching radius, as discussed in Section \ref{sec:mass_rate_derive}, only in a few local quasars can the neutral outflows extend to 15 kpc \citep{rubinKinematicsColdMetalenriched2022}, and the only spatially resolved observation of ISM \nai~in a $z=3$ massive quenched galaxy gives $r_{\rm out}=2.7$ kpc \citep{deugenioFastrotatorPoststarburstGalaxy2024}, consistent with our $r_{\rm out} = r_{\rm eff}$ assumption. We note an extreme case where a $z=2.42$ \nai~outflow is illuminated by a background quasar, extending 38 kpc away from its host galaxy \citep{morettiEmpiricalCalibrationNa2026}. In such a case, the fastest \DD\ outflow (the merging pair) can escape even the NFW potential. But it is yet unclear whether such extremely far outflows are common at high redshift. Thus, we argue that scenarios with $r_{\rm out}\lesssim7$ lend greater credence to the existence of fountain mechanisms in place already at $z\sim3.5$. In this case, the outflowing gas from predominantly quenched galaxies returns on relatively short timescales and is unlikely to be directly correlated with the most recent starburst that formed the bulk of stellar mass, as inferred from the SED modeling. The observed neutral outflows could just be tracing a feedback cycle typical of quenching maintenance mode in the \DD\ sample. In contrast, truly ejective events can occur during or after the main quenching epoch, in the presence of mergers or strong AGNs.

\section{AGN presence}
\label{sec: AGN}
\begin{figure}
    \centering
    \includegraphics[width=\linewidth]{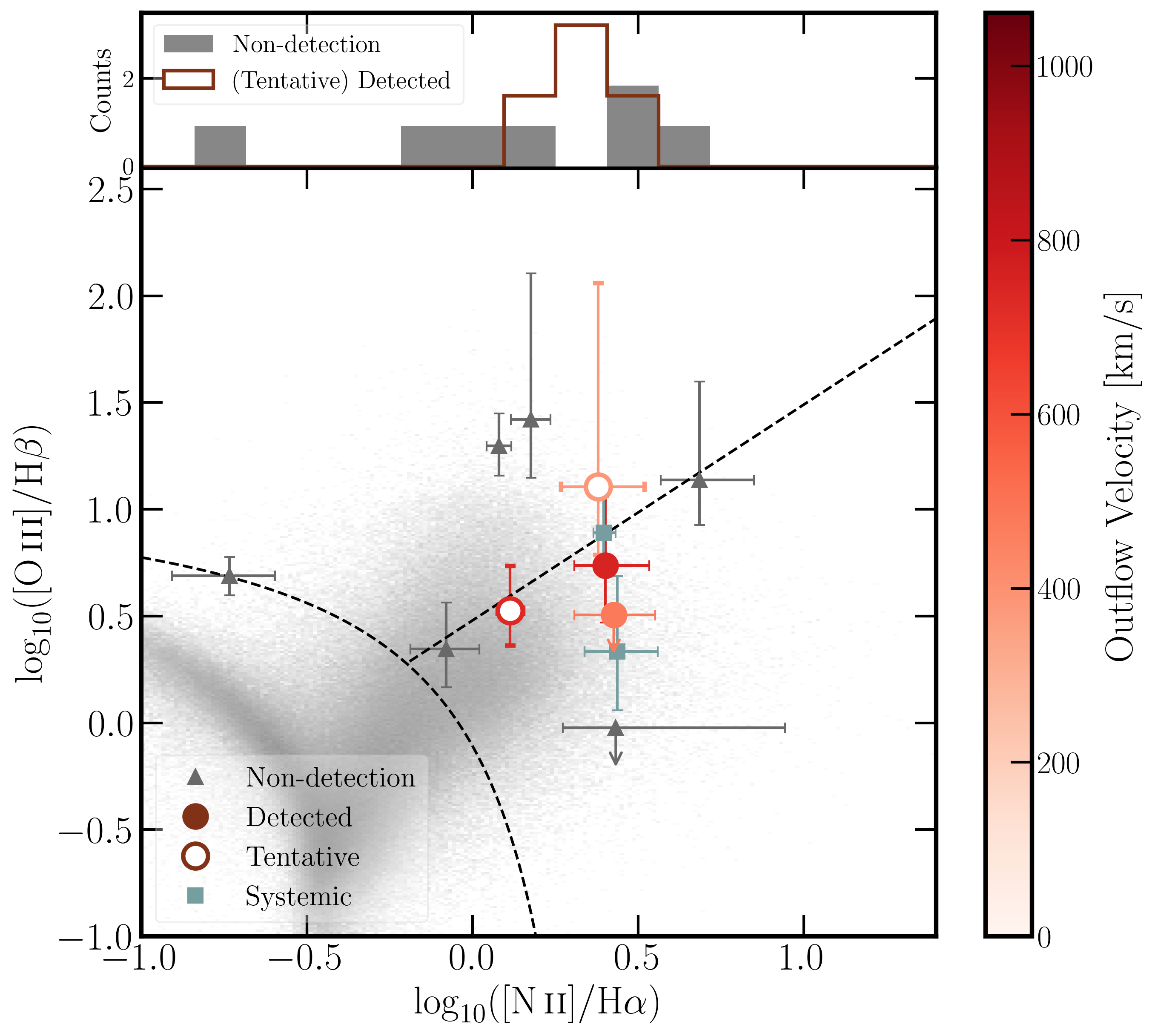}
    \caption{BPT diagram of the \DD~targets with a constrained \nii/\halpha\, ratio. The black dashed curve/line shows the $z=3$ \citet{kewleyTheoreticalEvolutionOptical2013}, and \citet{fernandesAlternativeDiagnosticDiagrams2010} demarcations of star-forming, AGN-dominated, and LI(N)ER-like galaxies, respectively. The black scatter in the background shows the SDSS DR7 galaxies \citep{abazajianSeventhDataRelease2009}. The blueshifted \nai~D (tentatively) detected DD targets are plotted as (hollow) filled circles, color-coded by their outflow velocity; the \nai~D systematic and non-detection targets are plotted as cyan squares and gray triangles, respectively. The histogram on top shows the normalized distribution of the \DD~targets along the \nii/\halpha\, axis, where grey represents the \nai~D undetected targets, and red represents the blueshift \nai~D (tentatively) detected targets.}
    \label{fig:BPT_v_out}
\end{figure}

In this section, we examine the possible AGN presence in the \DD\ galaxies. Fig. \ref{fig:BPT_v_out} shows the \DD\ targets plotted on the BPT diagnostic diagram \citep{baldwinCLASSIFICATIONPARAMETERSEMISSIONLINE1981}, color-coded by outflow velocity when available, based on the \halpha, \hbeta, \oiii\,$\lambda\,5007$, and \nii\,$\lambda\,6584$ emission lines, given that at least one group of emission lines is detectable.

In the \DD\ sample, 3/23 targets, DD-78, DD-111 and DD-229, exhibit broad emission-line features ($\sigma_{\rm H\alpha -broad}\sim3000-5000\,\,\rm km\,s^{-1}$), providing strong evidence of ongoing AGN activity. Moreover, DD-78, DD-111, DD-229, and DD-236, are the only sources with an X-ray detection in public catalogs \citep{itoDeepDiveDeepDive2025a}, with $L_{\rm2-10\,keV}>3\times10^{43}\,\mathrm{erg}\,\mathrm{s^{-1}}$, another clear sign of the presence of nuclear activity. Otherwise, most of the \DD\ targets, \nai~D detected or not, fall in the LI(N)ER part of the BPT diagram, or are undetected in the emission lines at all. We can conclude that approximately half of the galaxies in the \DD\ sample are LI(N)ER-like or host some level of AGN activities \citep[refer to][for local examples]{yanNatureLINERlikeEmission2012, belfioreSDSSIVMaNGA2016}.

However, we note that 9/13 of the \nai~D (tentatively) detected \DD\ galaxies (other than DD-185, DD-229, DD-236, and DD-270) do not show strong emission lines, nor are they X-ray confirmed. In fact, most of the \nai~D detected \DD\ targets show almost no signs of \hbeta\ and \oiii\ emission and exhibit net \halpha\ absorption \citep{itoDeepDiveDeepDive2025a}. Moreover, the broad-line, X-ray detected AGN targets DD-78 and DD-111 are \nai~D undetected. This suggests a lack of direct correlation between ongoing AGN activity and neutral outflows. However, this may be due to differences in timescales between observable outflows and AGNs. A similar scenario is proposed in \citet{taylorJWSTEXCELSSurvey2026}, where quasar mode AGN feedback in $z\sim3$ PSBs can be periodic and launch gas outflows on $\sim40$ Myr cycles. The only cases where clear, strong AGN evidence coincides with a strong neutral outflow are DD-229 (X-ray broad-line AGN with a possible P-Cygni \nai~D profile) and DD-236 (X-ray detected nuclear activity with an extreme excess \nai~D absorption feature and mass outflow rate), suggesting the most extreme outflows are indeed powered by active AGNs.

\section{Discussion}
\label{sec: discuss}

Our study of the 23 $z\sim3.5$ \DD\ quiescent galaxies indicates that neutral outflows that carry large gas masses are ubiquitous in massive quenched systems even at $z>3$. Neutral outflows are detected primarily in massive objects (see Fig. \ref{fig:z_vs_M} and Section \ref{sec: overview}), consistent with studies at similar redshifts. The highest-velocity outflows ($\gtrsim 750$ \kms) likely correlate with ongoing mergers or vigorous AGN activity and, in one case, exhibit a possible P-Cygni profile. Moreover, the outflow properties in our \DD\ sample are comparable with those of local counterparts (Fig. \ref{fig:mass_vel_sfr}, \ref{fig:age_voff}). For all targets, the mass outflow rates exceed the ongoing SFRs by at least 0.5 dex; however, because most of the outflows are unlikely to escape their hosts (Fig. \ref{fig:vesp_vout&sfr_mout} and Section \ref{sec: escape}), they could be recycled via fountain-like mechanisms with relatively short timescales. We still lack smoking-gun evidence to fully explain the driving mechanism of these neutral outflows and their role in quenching or maintaining quenching of the host galaxies; spatially resolved and high-resolution spectra are still needed.

\subsection{Comparison with the locals and redshift evolution}

Big sample studies of the local neutral outflows traced with either \mgii\ or \nai~D find increasing velocity offsets \voff\ with higher SFR \citep[e.g.,][]{concasTwofacesIonizedNeutral2019, davisExtendingDynamicRange2023, sunEvolutionGasFlows2024}. For PSBs, \voff\ are mostly in the range of roughly 0 to -200 \kms, with a median value of \voff$\sim0$ \kms~(Fig. \ref{fig:mass_vel_sfr} and \ref{fig:age_voff}). Yet in the $z\sim3.5$ \DD\ sample, we find no SFR-\voff\ correlation ($\rho_{\,\rm Spearman} = -0.6$, see Fig. \ref{fig:mass_vel_sfr}), similar to conclusions drawn from the $z\sim2$ sample \citetalias{daviesJWSTRevealsWidespread2024}. We note the possible SFR--$\dot{M}_{\rm out}$ correlation in the \DD\ sample ($\rho_{\,\rm Spearman} = 0.46$, see Fig. \ref{fig:vesp_vout&sfr_mout} and Section \ref{sec: mass outflow rate}), which is consistent with local Universe studies. However, the derivation of $\dot{M}_{\rm out}$ depends on many assumptions, and the correlation is not particularly strong; it is thus hard to draw any firm conclusions. The oldest quiescent SDSS galaxies are reported to show even low velocity net inflow \citepalias[i.e., positive \voff$\sim50$ \kms,][]{sunEvolutionGasFlows2024}. A few cases of inflow have also been reported in quenched systems at $z\sim2$ \citepalias{daviesJWSTRevealsWidespread2024}, in relatively old (quenched $\sim1$ Gyr ago) $z=2.7$ QGs \citep{bevacquaFeedingDeadNeutral2026}, as well as in $z\sim3$ PSBs \citep{taylorJWSTEXCELSSurvey2026}. We also find two targets with possible redshifted \nai~D features indicative of infalling gas, but the significance of \voff\ is currently marginal (see Table \ref{tab:outflow_properties} and Section \ref{sec: overview}).

One of the questions we are investigating is whether the properties of the neutral outflows evolve with redshift. There are no notable differences between the \DD\ sample, the $z\sim2$ \citetalias{daviesJWSTRevealsWidespread2024} sample, and the $z\sim0$ counterparts in \citetalias{sunEvolutionGasFlows2024}. As discussed in Section \ref{sec: compare} and shown in Fig. \ref{fig:age_voff}, on average, the \DD\ targets have \voff$\sim0\rm\,\, km\,s^{-1}$ that is very similar compared to local QGs \citep{sunEvolutionGasFlows2024} in the same mass and age bin. Our finding is consistent with a recent study of neutral outflows traced by \mgii\ in all publicly available $z=1-10$ \jwst/NIRSpec stacked spectra, which similarly shows no evidence for a $z-$\voff\ correlation \citep{lyuFirstStatisticalDetection2026}.

\subsection{Driving mechanisms}

\subsubsection{Mergers or star-formation as drivers}

The two targets (DD-185/196) with the highest outflow velocities (other than DD-229) are a merging pair identified by \citet{itoMergingPairMassive2025}. Their high sodium velocity offset is likely also attributable to ongoing interactions. The ``outflow" velocities of DD-185 and DD-196 are similar, as are the shapes of their excess \nai~D absorption profiles (Fig. \ref{fig:spec}), both are strongly blueshifted with broad and relatively weaker \nai~D profiles. Moreover, DD-134 is one member known as ``Jekyll'' in the interacting pair ``Jekyll and Hyde'' \citep{schreiberJekyllHydeQuiescence2018}. With spatially resolved data, \citet{perez-gonzalezAcceleratedQuenchingChemical2025} interpret the excess sodium absorption of DD-134 as a dark cloud extended from the companion galaxy ``Hyde'', covering DD-134 only in the foreground. For the other \DD\ targets, NIRCam images indicate potentially disturbed morphology in DD-179. \citet{baronNotWindyAll2024} find that for spatially resolved local PSBs, the observed blueshifted excess \nai~D absorption can very much be a result of early-stage (up to companions $\sim50$ kpc away) interactions rather than AGN winds. Thus, overdense nearby environments and gravitational interactions may partially be responsible for the removal of neutral gas in distant quenched systems. 

As concerns the role of possible recent star formation and the ensuing supernova feedback in launching massive neutral outflows, we find that the timescale for the outflows to rain back onto the galaxies (Section \ref{sec: escape}) is much shorter than the elapsed time since the last major starburst event. This suggests that the currently observed outflows are not likely to be directly related to the last starburst event. For the Keplerian potential, if $r_{\rm out}=$\reff, the return timescale is $\sim3$ Myr; if $r_{\rm out}=7$ kpc, the return timescale rises to $\sim60$ Myr. For the NFW potential, unless launching at $\gtrsim30$ kpc, the return timescale is $\sim110-180$ Myr. None of those return timescales is fully consistent with the time since the most recent starbursting event, which occurred $\sim400$ Myr ago \citep{hamadoucheDeepDiveTracingEarly2026}. We note that both \citetalias{daviesJWSTRevealsWidespread2024} and \citet{taylorJWSTEXCELSSurvey2026} have concluded that for $2\lesssim z\lesssim4.6$ galaxies, especially for QG/PSBs in them, the current star-formation or just supernova activities are not strong enough to power the outflows.

\subsubsection{AGN as drivers}
\label{sec: AGN as drivers}

A handful of $z>1$ studies link neutral outflows to (past) AGN feedback \citep[e.g.,][]{daviesNuclearCircumgalacticZooming2020, daviesJWSTRevealsWidespread2024, belliStarFormationShut2024, sunExtremeNeutralOutflow2026, taylorJWSTEXCELSSurvey2026}. In the local Universe, a large SDSS sample study by \citet{concasTwofacesIonizedNeutral2019} found no correlation between outflow strength and AGN occurrence in star-forming galaxies. Interestingly, all of the \nai~D (tentatively) detected \DD\ targets fall within the LI(N)ER region of the BPT diagram. While LINERs are often interpreted as low-luminosity AGNs, they do not necessarily host an active ``N-ucleus''; in such cases, they are referred to as LIERs. Indeed, studies have shown that in red galaxies, LI(N)ER emission is frequently powered by diffuse stellar sources (e.g., post-AGB stars; \citealt{yanNatureLINERlikeEmission2012, belfioreSDSSIVMaNGA2016}) rather than by AGN activity.
Moreover, the \nai~D undetected \DD\ targets show a similar distribution in the AGN/LI(N)ER region on the BPT diagram as the \nai~D detections. If we seek more substantial AGN evidence, most \nai~D detected \DD\ targets show no spectral signatures of broad-line AGN. Only 4/13 targets are X-ray-confirmed and/or host strong broad emission lines, of which two (DD-78 and DD-111) are undetected in \nai~D. The one target, DD-82, with very strong Balmer emission lines and a broad \halpha+\nii\ profile that we did not include in the sample (see Section \ref{sec: deepdive}), shows no excess \nai~D absorption features visually. We thus cannot establish a straightforward link between the presence of active AGN and blueshifted excess \nai~D absorption in the \DD\ sample. 

Two X-ray confirmed and/or broad-emission-line targets are detected strongly in excess \nai~D absorption: DD-229 shows a possible P-Cygni \nai~D profile, and DD-236 shows an extreme mass outflow rate. The P-Cygni profile in DD-229 typically relates to strong outflows \citep[e.g.,][]{prochaskaSimpleModelsMetalline2011}. It has the highest outflow velocity ($v_{\mathrm{out}}\sim750$ \kms) among all targets, other than the merging pair. DD-229 is similar to several spatially resolved local \nai~D P-Cygni profiles \citep[they are all strong, compact, and almost all links to ongoing broad-line AGNs, e.g.,][]{rupkeMultiphaseStructurePower2013, rupkeSpatiallyExtendedNa2015, pernaMultiphaseOutflowsMkn2019, baronMultiphaseOutflowsPoststarburst2020, baronNotWindyAll2024}. At higher redshift, DD-229 is similar to the case reported by \citet{deugenioFastrotatorPoststarburstGalaxy2024}, in which a spatially resolved AGN-driven neutral outflow in a $z=3$ massive QG is detected with high speed (\vout$\sim1000$ \kms, comparable to $v_{\mathrm{esc}}$) and at a small radius of 2.7 kpc. As for DD-236, it shows an extensive deep excess \nai~D absorption feature, yielding an EW of $\sim7\AA$. From the deep excess absorption feature in DD-236, we derived the most extreme \dMout$=2.68^{+0.27}_{-0.28}$ ever reported beyond the local Universe (Section \ref{sec: mass outflow rate}). It appears that the fastest and strongest outflows are indeed associated with ongoing AGN activity.

However, even in the absence of a direct relationship between active AGN presence and neutral outflow, either in literature or in our \DD\ sample, it is still very likely that the outflows we observed were driven by past AGN activities. For spatially resolved $z\sim2$ galaxies, \citet{daviesNuclearCircumgalacticZooming2020} find that the properties of AGN-induced ionized phase outflows are not related to the AGN luminosity, suggesting evolving AGN-outflows. \citetalias{daviesJWSTRevealsWidespread2024} proposes that the outflows in their $z\sim2$ sample are driven by the rapid blowout of past AGNs. Similarly, \citet{sunExtremeNeutralOutflow2026} determines the outflow in a $z=1.3$ QG with extreme \dMout$=2.40^{+0.11}_{-0.16}$ to be the relic of a recent AGN, and very recently, \citet{taylorJWSTEXCELSSurvey2026} proposes an episodic AGN scheme based on simulations, which expels neutral gas in $\sim$40 Myr cycles. We specifically note that the episodic AGN timescale proposed by \citet{taylorJWSTEXCELSSurvey2026} coincides with our estimation of the outflow return timescale ($\sim70$ Myr for $r_{\rm out}=7$ kpc and Keplerian potential, $\sim140$ Myr in NFW potential) to within an order of magnitude. We can thus argue that fossil/episodic AGNs once drove the neutral outflows and then (temporarily) shut down, explaining the lack of correlation between the presence of active AGNs and outflow velocity and strength. In any case, we cannot rule out that the \DD\ neutral outflows are AGN-driven, but they are at least not driven by ongoing AGN activity. Further spatially resolved study and larger sample statistics are required to understand whether the \DD\ outflows are relics of fossil/episodic AGN activity.

\subsection{Dynamic fountains}

Regardless of the driving mechanism, for the outflows to permanently quench their hosts by removing gas, they must at least escape their hosting galaxies. As discussed in Section \ref{sec: escape}, most \DD\ ``outflows" are not likely to escape their galaxies. We note that escapability depends on the assumed outflow radius and potentials. Assuming a Keplerian potential, the escape of outflows depends strongly on the assumed launch radius: (1) if launched at \reff\ ($\sim1$ kpc), most outflows cannot escape even the baryon-dominated inner region ($<2$\reff); (2) if launched at $\sim7$ kpc, roughly half may escape; and (3) if launched at $\sim15$ kpc, most can escape. In contrast, assuming an NFW potential, outflows generally remain bound unless launched from radii $\gtrsim30$ kpc. We argued in Section \ref{sec: mass outflow rate} and \ref{sec: escape} that it is unlikely that the \DD\ targets host extremely far-away outflows (i.e., $\gtrsim15$ kpc), as the only spatially resolved example of a massive $z=3$ PSB has \nai~traced outflow at 2.7 kpc \citep{deugenioFastrotatorPoststarburstGalaxy2024}, comparable to the \DD\ effective radius. Even assuming a relatively large 7 kpc outflow radius, at least half of the \DD\ outflows cannot escape their hosts. The evidence suggests that a fountain-like mechanism recycles the outflows, which are part of the galactic baryon cycle. Although the \DD\ neutral outflows carry significant mass (Fig. \ref{fig:vesp_vout&sfr_mout}), because they do not effectively remove the gas, it is questionable whether these outflows directly contribute to the permanent quenching of galaxies, yet they may play a role in the feedback-regulated quenching maintenance processes.

Instead of being removed, we propose that the neutral outflows cycle in dynamic fountains on relatively short timescales, again depending on the assumed launching radius and potential. In either the Keplerian or NFW potential, the bound outflow recycles on timescales of $10^{10}$ Myr or less (see Section \ref{sec: escape}). As discussed in Section \ref{sec: AGN as drivers}, our outflow fountain timescale estimate is consistent with the simulation-based episodic AGN timescale proposed by \citet{taylorJWSTEXCELSSurvey2026}. It is possible that those quick dynamical fountains are powered by episodic AGNs, which would reignite the AGNs \citep{taylorJWSTEXCELSSurvey2026}; and, as the \texttt{Magneticum Pathfinder} simulation predicts for massive QGs at $z>3$, boost minor outskirt star-formation rejuvenation \citep{remusRelightCandleWhat2025} when they cycle back.

In the analysis, we have treated the two redshifted \nai~D-detected targets as having ``systemic'' velocities, given their low velocity offsets and potential geometric and resolution effects. However, we note that both local \citepalias[e.g.,][]{sunEvolutionGasFlows2024} and $z>2$ studies \citep{daviesJWSTRevealsWidespread2024, bevacquaFeedingDeadNeutral2026, taylorJWSTEXCELSSurvey2026} have reported a small number of redshifted \nai~D detections, with detection rates consistent with our findings ($\sim10\%$). If those redshifted \nai~D detections are indeed signatures of inflowing gas, it is possible that they trace the backflow component of the dynamic fountain.

In any case, we expect the neutral outflows in the \DD\ sample to recycle on relatively short timescales, and thus are unlikely to play a major role in quenching their hosts, but may play a role in the feedback-dominated quiescent maintenance. The outflows are unlikely to be directly related to ongoing AGN activity or the most recent starbursts, but may be powered by fossil/episodic AGN activity and reignite AGN or boost outskirt minor star formation as they cycle back.

\subsection{Caveats}

Finally, we note some caveats of this work. 1) We only (tentatively) detected $\Delta v \lesssim -150$ \kms, which might be partly affected by the limit of NIRSpec grism's medium resolution ($R\sim1000$, which corresponds to $\sim100$ \kms), which limits sensitivity to smaller velocity shifts. This is why we included the \nai~D systemic and low-\voff\ redshifted targets in the analysis, as they may exhibit resolution-limited outflows. 2) The detection of excess \nai~D absorption heavily depends on the stellar continuum modeling, since the ISM \nai~D component is only measurable on the continuum-normalized spectra. For $z\sim2$ galaxies spanning $8.5<\log(M_\star/M_\odot)<11.7$ and $-13<\log(\mathrm{sSFR\,[yr^{-1}]})<-7$, \citetalias{daviesJWSTRevealsWidespread2024} tested stellar absorption profiles for a range of sodium abundances, and found that extremely sodium-enhanced stellar populations ([Na/Fe] = 0.6) can produce extra absorption with an EW up to 1.2 \AA. This value is comparable to the excess \nai~D EW (measured on the continuum-normalized spectra) of the tentative detections DD-96 and DD-170, but weaker than all the other observed excess absorption profiles with excess \nai~D EW of 1.4--7.0 \AA\ (see Table \ref{tab:outflow_properties}). 3) Estimating $N(\hi)$ and especially converting excess \nai~D absorption to mass outflow rates requires many assumptions, largely calibrated on local observations (see Section \ref{sec:mass_rate_derive}). \nai~D is also not the most ideal tracer for neutral gas, but beyond the local universe, \hi\ 21cm hyperfine line is extremely hard to observe. Moreover, studies have shown that molecular outflows carry a substantial mass budget \citep[e.g.,][]{ciconeMassiveMolecularOutflows2014,herrera-camusMolecularIonizedGas2019, kimFirstResultsSMAUG2020}, but they are very difficult to detect \citep[see][and references therein]{veilleuxCoolOutflowsGalaxies2020}. 4) Local studies have found strong dependence of \nai~D traced neutral outflow detection on inclination (or geometry, in a more general term): \citet{concasTwofacesIonizedNeutral2019} found a clear dependence of \voff\ on inclination for SDSS galaxies: face-on targets show, on average, higher \voff\ than edge-on galaxies. \citetalias{sunEvolutionGasFlows2024} also found a similar inclination-\voff\ relation for a larger SDSS sample, as well as a decreasing outflow detection fraction for higher inclinations. Even though we do not have much inclination difference for the compact elliptical galaxies in \DD, geometric effects could still bias the measured EW and \voff\ values.

\section{Summary}
\label{sec: summary}

We present a systematic census of \nai-traced neutral outflows in a sample of 23 massive quiescent galaxies at $2.8<z<4.6$ from the \DD\ program. Our main findings are:

\begin{itemize}

\item Incidence of neutral outflows:
Excess \nai~D absorption is (tentatively) detected in 13/23 galaxies (57\%), with 7/23 (30\%) showing blueshifted absorption indicative of outflows. All detections occur in the massive regime ($\log M_\star/M_\odot>10.5$), consistent with trends seen at lower redshift.

\item Outflow kinematics:
Blueshifted systems show velocity offsets of $\Delta v\lesssim -150$~km\,s$^{-1}$, with three reaching $\Delta v \sim -500$~km\,s$^{-1}$. The fastest outflows are associated with either an ongoing merger (DD-185/196) or AGN activity with a possible P-Cygni signature (DD-229); the most massive outflow (DD-236, \dMout$\approx 2.7$) is also associated with an X-ray AGN. The \DD\ outflow velocities overlap with those measured in local PSBs when matched in stellar mass and post-burst age. The detection fraction is slightly higher than in local and $z\sim2$ samples, though this may partly reflect mass and S/N differences.

\item Escapability and recycling:
Using dynamical masses and effective radii, most outflows have $v_{\rm out}<v_{\rm esc}$, implying bound orbits and short fountain return times ($\sim3-180$ Myr depending on $r_{\rm out}$ and potential profile). Even assuming larger launch radii, half of the outflows remain bound. Therefore, these winds are likely recycling rather than evacuating the galaxy.

\item Connection to quenching:
All \nai-detected galaxies exhibit mass outflow rates exceeding their current SFRs, in several cases by more than an order of magnitude. One source (DD-236) shows \dMout$\approx 2.7$, the most extreme neutral outflow rate reported beyond the local Universe. But since most outflows are not likely to escape, their direct role in quenching or maintaining quiescence is uncertain.

\item Driving mechanisms:
\DD\ galaxies' short return timescales ($\sim$3–180 Myr) differ from the $\sim$400 Myr since the last starburst, suggesting they are likely not starburst-driven winds. Most \nai-detected \DD\ galaxies lie in the LI(N)ER region of the BPT diagram, and so are the non-detections, and only 2/13 detections are X-ray confirmed. Two broad-line X-ray AGNs are not \nai~detected. We do not establish a clear AGN-outflow correlation. However, we note that the outflows may still be driven by past AGN activity, and the absence of a direct correlation may be due to differences in the timescales of the involved processes. We also note that the strongest outflows (AGN-driven DD-229/236 and the merging pair DD-185/196) demonstrate that both nuclear activity and interactions can enhance neutral winds. 

\item Overall picture:
Neutral outflows in massive quiescent galaxies at $z>3$ are common, similar to low-$z$ counterparts, and carry substantial mass, yet usually remain gravitationally bound. They likely represent short-lived fountain cycles rather than ejective quenching events. Their origin is not directly tied to ongoing AGN activity or the most recent starbursts, but may reflect a mix of past AGN activity and ongoing mergers.

\end{itemize}

\begin{acknowledgements}
The authors sincerely appreciate the valuable suggestions and comments by Alice Concas, Stefano Zibetti, Laura Scholz-Diaz, and Sirio Belli. The authors especially thank Filippo Fraternali for inspiring the toy model for the outflow return timescale and Cecilia Bacchini for help with \texttt{galpy}. KI, FV, and PZ acknowledge support from the Independent Research Fund Denmark (DFF) under grant 3120-00043B. TK acknowledges support from JSPS grant 25KJ1331. MO acknowledges support from  JSPS KAKENHI Grant Number JP25K07361. WMB gratefully acknowledges support from DARK via the DARK fellowship. This work was supported by a research grant (VIL54489) from VILLUM FONDEN. This work is based on observations made with the NASA/ESA/CSA James Webb Space Telescope. The data were obtained from the Mikulski Archive for Space Telescopes at the Space Telescope Science Institute, which is operated by the Association of Universities for Research in Astronomy, Inc., under NASA contract NAS 5-03127 for JWST. These observations are associated with program \#3567. The data described here may be obtained from
\url{https://dx.doi.org/10.17909/T9RP4V}. Some of the data products presented herein were retrieved from the Dawn JWST Archive (DJA), DJA is an initiative of the Cosmic Dawn Center (DAWN), which is funded by the Danish National Research Foundation under grant DNRF140. 
\end{acknowledgements}

\bibliographystyle{aa.bst}
\bibliography{outflows.bib}

\end{document}